\documentclass[aps,prd,amsmath,amsfonts,a4paper,11pt,reprint,onecolumn,notitlepage,square,numbers,nofootinbib,sort&compress]{revtex4-1}
\usepackage{epsfig}
\usepackage[english]{babel}
\usepackage{amsfonts}
\usepackage{amssymb}
\usepackage{amsmath}
\usepackage{booktabs}
\usepackage{colortbl}
\usepackage{multirow}
\usepackage{bm}
\usepackage{hyperref}
\usepackage[labelfont={bf,scriptsize},textfont={scriptsize},justification=RaggedRight,format=hang]{caption,subfig}
\usepackage{natbib}

\def\d{\delta}
\def\e{\epsilon}

\def\={\nonumber &=}

\def\({\left(}
\def\){\right)}
\def\[{\left[}
\def\]{\right]}
\def\<{\left\langle}
\def\>{\right\rangle}

\def\bn{{\bf n}}

\def\bx{{\bf x}}

\def\curl{\mathcal}

\def\eq{\begin{align}}
\def\qe{\end{align}}
\def\eqa{\begin{eqnarray}}
\def\qea{\end{eqnarray}}

\def\and{\quad \mbox{and} \quad}

\def\fnl{f_\mathrm{NL}}

\def\fnlhat{\hat{f}_{\mathrm{NL}}}
\def\Fnl{F_\mathrm{NL}}
\def\Fnlhat{\hat{F}_{\mathrm{NL}}}

\def\fnllocal{f_\mathrm{NL}^\text{loc}}

\def\fnlequil{f_\mathrm{NL}^\text{eq}}

\def\fnlequil{f_\mathrm{NL}^\text{eq}}

\def\bfnl{\kern2pt\overline{\kern-2ptf}_\mathrm{NL}}

\def\lmax{l_\text{max}}

\def\kmax{k_\text{max}}

\def\barQ{\kern2pt\overline{\kern-2pt\curl{Q}}}

\def\barR{\kern2pt\overline{\kern-2pt\curl{R}}}

\def\prd{Physical Review D}
\def\mnras{Monthly Notices of the Royal Astronomical Society}

\def\apj{Astrophysical Journal}
\def\apjs{Astrophysical Journal, Supplement}

\newcommand{\para}[1]{\par\vspace{2mm}\noindent\textbf{{#1}}}
\newcommand{\vect}[1]{\bm{\mathrm{{#1}}}}
\renewcommand{\d}{\mathrm{d}}
\renewcommand{\e}[1]{\mathrm{e}^{{#1}}}
\newcommand{\im}{\mathrm{i}}
\renewcommand{\leq}{\leqslant}

\newcommand{\Nscal}{N_{\text{scal}}}
\newcommand{\Nmax}{N_{\text{max}}}

\newcommand{\ipleft}{\langle\kern-2.5pt\langle}
\newcommand{\ipright}{\rangle\kern-2.5pt\rangle}

\newcommand{\baralpha}{\bar{\alpha}}
\newcommand{\barbeta}{\bar{\beta}}
\newcommand{\barN}{\bar{N}}
\newcommand{\bargamma}{\bar{\gamma}}

\DeclareMathOperator{\Or}{\mathcal{O}}

\newcolumntype{s}{>{\columncolor[gray]{0.9}$\displaystyle}c<{$}}
\newcolumntype{S}{>{$\displaystyle}c<{$}}

\begin{document}

%%%%%%%%%%%%%%%%%%%%%%%%%%%%%%%%% INTRODUCTION %%%%%%%%%%%%%%%%%%%%%%%%%%%%%%%%

\title{General CMB bispectrum analysis using wavelets and separable modes}

\author{Donough Regan}
\affiliation
{Astronomy Centre, University of Sussex, Brighton BN1 9QH, United Kingdom}

\author{Pia Mukherjee}
\affiliation
{Astronomy Centre, University of Sussex, Brighton BN1 9QH, United Kingdom}

\author{David Seery}
\affiliation
{Astronomy Centre, University of Sussex, Brighton BN1 9QH, United Kingdom}

\begin{abstract}
	In this paper we combine partial-wave (`modal')
	methods with a wavelet analysis
	of the CMB bispectrum.
	Our implementation exploits the advantages of
	both approaches
	to produce
	robust, reliable and efficient estimators
	which can constrain the amplitude of
	arbitrary primordial bispectra.
	This will be particularly important for upcoming surveys
	such as \emph{Planck}.
	A key advantage is the computational efficiency of
	calculating the inverse covariance matrix in wavelet space,
	producing an error bar which is close to
	optimal.
	We verify the efficacy and robustness of the method by
	applying it to WMAP7 data, finding
	$\fnllocal=38.4\pm 23.6$
	and $\fnlequil=-119.2\pm123.6$.
\end{abstract}

\maketitle

%%%%%%%%%%%%%%%%%%%%%%%%%%%%%%%%% INTRODUCTION %%%%%%%%%%%%%%%%%%%%%%%%%%%%%%%%

\section{Introduction}
\label{sec:introduction}
In inflationary scenarios,
any measurable deviation
of the primordial density fluctuation
from Gaussianity
offers a window onto
the underlying physics.
Observable deviations
require violation~\cite{Maldacena2003} of at
least one assumption of the simplest inflationary model:
(i) a single, canonically normalized light degree of freedom;
(ii) slow-roll dynamics, and
(iii) the Bunch--Davies vacuum state on deep subhorizon scales.%
	\footnote{Some non-Gaussianity is always produced by post-inflationary
	gravitational reprocessing of the density fluctuation.
	In Einstein gravity this is expected to produce a signal of
	order $\fnl \sim \Or(1)$, but may be larger
	in modified theories of gravity \cite{10033270}.
	}
For comprehensive reviews see, eg., \cite{0406398,10036097,Chen2010}.

The prospect of recovering definite information about
inflationary microphysics has made the study of non-Gaussianity
an area of active research.
Much of this activity has been stimulated by
maps
produced by all-sky CMB experiments,
including the Wilkinson Microwave Anisotropy Probe
(WMAP) \cite{WMAP7}.
To maximize the value of these datasets requires
%general, accurate and robust
methods
capable of extracting
the primordial non-Gaussian signal.
The amplitude of this signal
is conventionally expressed as
$\fnl$.

An optimal bispectrum estimator
for $\fnl$
was developed in Ref.~\cite{Babich05}
and implemented by Smith et al.~\cite{SmithetAl2009}.
For primordial non-Gaussianity in the local mode
(to be defined in Section~\ref{sec:modesreview} below)
it gives the constraint $\fnl = 38 \pm 21$ from 5-year WMAP data.
Unfortunately,
optimality of the method requires calculation
and inversion of a pixel-by-pixel
covariance matrix. Since WMAP maps contain $\Or(10^6)$
pixels this is an onerous task.
The calculation can be simplified by approximating the covariance
matrix as diagonal, but
the effects of anisotropic noise and masking
make
this approximation
degrade with increasing resolution.
The diagonal approximation is unlikely to be satisfactory for \emph{Planck}.

Alternative approaches exist, which aim to reduce the calculational
burden at the expense of a sub-optimal error bar.
Fergusson, Liguori \& Shellard suggested that
the bispectrum could be decomposed into partial
waves or `modes'~\cite{FLS1}.
In this approach,
an efficient inverse covariance weighting has been proposed
which renders the estimator closer to optimal~\cite{FS3}.
But a potential drawback is the necessity to
remove cross-terms (mainly generated by anisotropic noise)
by subtracting a linear
correction.
For an experiment such as \emph{Planck}
the required cancellation may be very precise
if the diagonal approximation is used for the inverse covariance
matrix.
A preliminary application of the method has reduced the error bar
from $\Delta \fnl = 29.5$ to $\Delta \fnl=27.6$~\cite{FLS10}.
%making
%progress towards the optimal result (for data out to $\lmax=500$) $\Delta \fnl=22.9$
%\cite{FLS10}.

Instead of partial waves
one can use wavelets and needlets,
which have a long history of
application to CMB analysis
\cite{Cayon,MukWang04,Baldi,Pietrobon,Lan}.
Like a decomposition into partial waves,
the advantage of these methods
is compression of the WMAP data from $\Or(10^6)$ pixels
into $\Or(10^3)$ wavelet coefficients, for which
calculation of the inverse covariance matrix
is comparatively trivial.
Despite this, the error bars achieved by this method
are only marginally worse than those produced by the
pixel-by-pixel approach.
Further,
Donzelli et al.~\cite{Donzelli2012}
have shown that, for wavelets, the linear correction term described above
is not necessary because scale-by-scale subtraction of the
mean for each coefficient
produces effective decorrelation
(see also Ref.~\cite{Curto2011}).
Even in the absence of these desirable properties,
wavelet approaches would be interesting because
the use of complementary methods
helps establish
sensitivities to contaminants such as foreground noise
and masking.
For example,
they have proven useful in performing an exacting noise analysis
\cite{Curto0807}.

These properties motivate
wavelet approaches to the CMB bispectrum.
However, we usually wish to use the CMB to constrain models of
early-universe inflation.
These do not generate predictions
for the CMB directly, and therefore
it is not simple to compare them to wavelet
coefficients recovered from the CMB: they predict
the correlation functions of the \emph{curvature perturbation},
$\zeta(\vect{k})$,
which can be translated
into the gravitational potential
$\Phi(\vect{k})$.
The most observationally important
of these are the two- and three-point functions,
$\langle \Phi(\vect{k}_1) \Phi(\vect{k}_2) \rangle$
and
$\langle \Phi(\vect{k}_1) \Phi(\vect{k}_2) \Phi(\vect{k}_3) \rangle$,
and
it is the parameters of these correlation functions
which we wish to estimate from the data.
But to do so they
must be converted into predictions for
statistical properties of the CMB anisotropies.

Currently,
the only practical way to carry out this conversion is to
systematically approximate each correlation function
using the
partial-wave expansion suggested by
Fergusson et al.~\cite{FLS1}, described above.
As we will explain in Section~\ref{subsec:primdecomp},
this has the effect of rendering
the calculation numerically tractable.
Funakoshi \& Renaux-Petel~\cite{Funak2012}
have recently detailed a formalism
to compute a suitable expansion directly from the rules of the
Schwinger (or `in--in') formulation of quantum field theory.
Once this has been accomplished it is straightforward to
write an estimator for each parameter appearing in
$\langle \Phi(\vect{k}_1) \Phi(\vect{k}_2) \Phi(\vect{k}_3) \rangle$.
We will review some aspects of the modal decomposition in
Section~\ref{sec:modesreview},
and explain how suitable estimators can be constructed.
For further details we refer to the
literature~\cite{FLS1,FLS10,RSF10,FRS2,FergussonShellard2007,FS2}.

In this paper we take the logical step of combining a
modal decomposition of each primordial $n$-point function
with the use of wavelet-based CMB estimators.
The CMB analysis is performed in wavelet-space
and takes advantage of an inverse
covariance matrix which is numerically much less challenging
than the pixel-by-pixel case.
A change-of-basis matrix allows us to map from
wavelet-space to modal-space. This approach
exploits the computational benefits of a wavelet-based estimator
but simultaneously allows comparison to the predictions of
primordial inflationary models.

\para{Summary.}
In Section~\ref{sec:modesreview} we explain the
partial-wave or `modal'
expansion technique and review its use in
CMB bispectrum analysis, and
in Section~\ref{sec:waveletsreview} we
summarize the methodology of wavelet-based estimators.
In Section~\ref{sec:primtolate} we describe a
prescription for projecting directly from the modal expansion of an
arbitrary
primordial bispectrum to the CMB bispectrum.
We also review the use of the modal expansion to
create simulated maps.

In Section~\ref{sec:waveletsmodes} we begin the application of
partial-wave expansions to the
wavelet-based estimators of Section~\ref{sec:waveletsreview}.
In Section~\ref{sec:wmaptest} we describe
an implementation of this prescription for the 7-year WMAP
data up
to $\lmax = 1000$.
Constraints on the amplitude of the constant, local, equilateral,
flattened and orthogonal models are given in
Section~\ref{sec:wmapconstraints}. We conclude
in Section~\ref{sec:conclusions}.

Sections~\ref{sec:modesreview}--\ref{sec:primtolate}
are a summary of the literature, and have been included to
fix notation and
make our presentation self-contained.
Readers familiar with the `modal' methodology and
wavelet-based CMB analysis may wish to proceed directly to
%Section~\ref{sec:wmaptest},
Section~\ref{sec:waveletsmodes},
making use of references to preceding sections where necessary.

\section{CMB bispectrum and partial-wave techniques}\label{sec:modesreview}

\subsection{CMB bispectrum}

In linear perturbation theory,
the spherical harmonic transform of the CMB temperature map
$\Delta T(\hat{\vect{n}})/T$
may be expressed in terms of the primordial gravitational
potential $\Phi$,
\begin{align}\label{almphi}
	a_{l m} = 4\pi (-\im)^l
	\int \frac{\d^3 k}{(2\pi)^3} \Delta_l (k) \Phi(\vect{k})
	Y_{l m}^*(\hat{\vect{k}})\,,
\end{align}
where
the unit vector $\hat{\vect{n}}$ determines an
orientation on the sky, and
$\Delta T(\hat{\vect{n}})/T = \sum_{ lm }a_{lm}Y_{lm}(\hat{\vect{n}})$.
In what follows we work up to $\lmax = 1000$.
The $\Delta_l (k)$ are transfer functions,
which map from primordial times to the surface of last scattering.
They are computed by solving the collisional Boltzmann equations
using publicly-available codes
such as CAMB~\cite{9911177} and CLASS~\cite{CLASS}.

\para{Angular bispectrum.}
The CMB bispectrum
is defined as the three-point correlation function of the $a_{l m}$,
namely
$B^{l_1 l_2 l_3}_{m_1 m_2 m_3} \equiv
\langle a_{l_1 m_1}a_{l_2 m_2}a_{l_3 m_3} \rangle$.
It can be written
\begin{align}\label{bispLs}
	B^{l_1 l_2 l_3}_{m_1 m_2 m_3} =
	(4\pi)^3 (-\im)^{l_1 +l_2 +l_3}
	\int \bigg(
		\prod_{i=1}^3 \frac{\d^3 k_i}{(2\pi)^3}
		\Delta_{l_i}(k_i)
		Y^*_{l_i m_i}(\hat{\vect{k}}_i)
	\bigg)
	\langle \Phi(\vect{k}_1)\Phi(\vect{k}_2)\Phi(\vect{k}_3) \rangle
	\,.
\end{align}
We recall that the primordial two- and three-point functions satisfy
\begin{align}
	\langle
		\Phi(\vect{k}_1) \Phi(\vect{k}_2)
	\rangle
	& = (2\pi)^3 \delta(\vect{k}_1 + \vect{k}_2)
	P_{\Phi}(k_1)
	\label{eq:p-def}
	\\
	\langle
		\Phi(\vect{k}_1) \Phi(\vect{k}_2) \Phi(\vect{k}_3)
	\rangle
	& = (2\pi)^3 \delta(\vect{k}_1+\vect{k}_2+\vect{k}_3)
	B_{\Phi}(k_1,k_2,k_3)\,.
	\label{eq:b-def}
\end{align}
Each inflationary model predicts
a specific form for $P_{\Phi}$ and $B_{\Phi}$.
A typical model will predict the appearance of
a finite number of
momentum-dependent combinations (or `shapes') in
$B_{\Phi}$, with amplitudes that depend on parameters of
the model.
These shapes can be regarded as similar to the
different Mandelstam channels in $2 \rightarrow 2'$
scattering, or
the structure functions of the hadronic tensor
$W^{\mu\nu}$ in QCD.
By constructing an estimator for the amplitude of each
shape we obtain observational constraints on
these parameters,
and in some cases it may even be possible to
rule out a model entirely.
However,
before any comparison with
observation we must first translate
Eqs.~\eqref{eq:p-def}--\eqref{eq:b-def}
into statistical properties of the
CMB anisotropy.

The $\delta$-functions in \eqref{eq:p-def}--\eqref{eq:b-def}
enforce momentum conservation.
For the bispectrum this requires that the momenta form a closed
triangle,
and 
implies that $B_{\Phi}$
may be expressed as a function of the $k_i$
alone.\label{pageref:k-magnitudes}
To express the $\delta$-function in multipole space, we use the
identity
\begin{equation}\label{deltaexp}
	\delta(\mathbf{k}_1+\mathbf{k}_2+\mathbf{k}_3)
	=
	8 \sum_{l_i m_i} \im^{l_1+l_2+l_3}
	\int \d x \; x^2
	\bigg(
		\prod_{i=1}^3 j_{l_i}(k_i x)Y_{l_i m_i}(\hat{\vect{k}}_i)
	\bigg)
	\int \d \Omega(\hat{\vect{x}}) \;
	\prod_{i=1}^3 Y^*_{l_i m_i}(\hat{\vect{x}})
	\,,
\end{equation}
where $j_n(x)$ is
a spherical Bessel function
and $\d \Omega$ is an element of area on the sphere.
After substitution into~\eqref{bispLs} we conclude
\begin{equation}
\begin{split}
	B^{l_1 l_2 l_3}_{m_1 m_2 m_3}
	=
	\left( \frac{2}{\pi} \right)^3
	\int \d x \, & \d k_1 \, \d k_2 \, \d k_3 \;
	(x k_1 k_2 k_3)^2
	\Delta_{l_1 }(k_1)
	\Delta_{l_2 }(k_2)
	\Delta_{l_3 }(k_3)
	B_{\Phi}(k_1,k_2,k_3)
	\\
	& \mbox{} \times
	j_{l_1}(k_1 x) j_{l_2}(k_2 x) j_{l_3}(k_3 x)
	\int \d\Omega(\hat{\vect{x}}) \;
	Y^*_{l_1 m_1}(\hat{\vect{x}})
	Y^*_{l_2 m_2}(\hat{\vect{x}})
	Y^*_{l_3 m_3}(\hat{\vect{x}})\,.
\end{split}
\label{eq:bispectrum-interim}
\end{equation}
To simplify~\eqref{eq:bispectrum-interim},
we note that the Gaunt integral is defined by
\begin{equation}\label{Gauntexp}
	\mathcal{G}^{l_1 l_2 l_3}_{m_1 m_2 m_3}
	\equiv
	\int \d\Omega(\hat{\vect{x}}) \;
	Y^*_{l_1 m_1}(\hat{\vect{x}})
	Y^*_{l_2 m_2}(\hat{\vect{x}})
	Y^*_{l_3 m_3}(\hat{\vect{x}})
	=
	h_{l_1 l_2 l_3}
	\bigg( \begin{array}{ccc}
		l_1 & l_2 & l_3 \\
		m_1 & m_2 & m_3 \end{array}
	\bigg)\,,
\end{equation}
where
$
\bigg( \begin{array}{ccc}
l_1 & l_2 & l_3 \\
m_1 & m_2 & m_3 \end{array} \bigg)
$
denotes the Wigner $3$-j symbol, and 
$h_{l_1 l_2 l_3} \equiv \sqrt{\dfrac{(2l_1+1)(2l_2+1)(2l_3+1)}{4\pi}}\bigg( \begin{array}{ccc}
l_1 & l_2 & l_3 \\
0 & 0 & 0 \end{array} \bigg)$.
The Gaunt integral is the analog of the Dirac $\delta$-function in
multipole space, and imposes constraints on the $l_i$.
Finally, we define the reduced bispectrum, $b_{l_1 l_2 l_3}$, to satisfy
\begin{equation}
	B^{l_1 l_2 l_3}_{m_1 m_2 m_3}
	=
	\mathcal{G}^{l_1 l_2 l_3}_{m_1 m_2 m_3} b_{l_1 l_2 l_3}\,,
	\label{eq:reduced-bispectrum}
\end{equation}
and it follows that
\begin{equation}
	b_{l_1 l_2 l_3}
 	=
 	\left( \frac{2}{\pi} \right)^3
 	\int \d k_1 \, \d k_2 \, \d k_3 \;
 	(k_1 k_2 k_3)^2
 	\Delta_{l_1 }(k_1)
 	\Delta_{l_2 }(k_2)
 	\Delta_{l_3 }(k_3)
 	B_{\Phi}(k_1,k_2,k_3)
 	\int \d x \; x^2 j_{l_1}(k_1 x) j_{l_2}(k_2 x)j_{l_3}(k_3 x)\,.
\end{equation}

\para{Shape function.}
We define the `local' bispectrum to satisfy
\begin{equation}\label{eq:locmodel}
	B_{\Phi}^{\text{loc}}(k_1,k_2,k_3)
	= 2 \Big(
		P_{\Phi}(k_1) P_{\Phi}(k_2)
		+ P_{\Phi}(k_1) P_{\Phi}(k_3)
		+ P_{\Phi}(k_2) P_{\Phi}(k_3)
	\Big)\,.
\end{equation}
For any bispectrum we can define a dimensionless `shape' function
by the rule
\begin{equation}\label{eq:shapeloc}
	S_{\Phi}^{\text{(loc)}}(k_1,k_2,k_3)
	\equiv
	\frac{B_{\Phi}(k_1,k_2,k_3)}{B_{\Phi}^{\text{loc}}(k_1,k_2,k_3)}\,.
\end{equation}
This choice is arbitrary: we could equally well have
defined a shape function by comparison to a
fiducial bispectrum other than $B_{\Phi}^{\text{loc}}$.
In the modal decomposition literature, the choice
$(k_1 k_2 k_3)^2 B_{\Phi}(k_1,k_2,k_3)$ is often made.
In this paper we adopt~\eqref{eq:shapeloc}
for numerical purposes, because it often proves
more stable.
To clearly distinguish our choice when comparing with the literature
we also define a canonical shape function $S_{\Phi}$,
\begin{align}
    \label{eq:shapecanon}
	S_{\Phi}(k_1,k_2,k_3)
	=
	\frac{(k_1 k_2 k_3)^2}{N} B_{\Phi}(k_1,k_2,k_3)\,,
\end{align}
where the normalization constant $N$ is adjusted
to ensure $S_{\Phi}(k,k,k)=1$.
With these choices
the reduced bispectrum may be written
\begin{equation}\label{redBisp}
\begin{split}
	b_{l_1 l_2 l_3}
	= 6
	\left( \frac{2}{\pi} \right)^3
	\int \d k_1 \, & \d k_2 \, \d k_3 \;
	(k_1 k_2 k_3)^2
	\Delta_{l_1 }(k_1)
	\Delta_{l_2 }(k_2)
	\Delta_{l_3 }(k_3)
	P_{\Phi}(k_1) P_{\Phi}(k_2)
	S_{\Phi}^{\text{(loc)}}(k_1,k_2,k_3)
	\\
 	& \mbox{}
 	\times \int \d x \; x^2 j_{l_1}(k_1 x) j_{l_2}(k_2 x)j_{l_3}(k_3 x)\,.
\end{split}
\end{equation}

\subsection{Primordial decomposition}\label{subsec:primdecomp}

Eq.~\eqref{redBisp} shows that conversion of the primordial
two- and three-point functions into predictions
for the statistical properties of the CMB
is simplified
whenever the shape function is \emph{separable}, ie., of the form
$S^{\text{(loc)}}_{\Phi} = X(k_1)Y(k_2)Z(k_3) + \text{perms}$.
In such cases the $k_1$, $k_2$ and
$k_3$ integrals in Eq.~\eqref{redBisp} can be decoupled,
greatly reducing the computational time.

To take advantage of this simplification,
Fergusson, Liguori \& Shellard suggested that an arbitrary
(not necessarily separable) shape function $S_{\Phi}^{\text{(loc)}}$
could be decomposed into a basis of separable partial waves~\cite{FLS1}.
The precise choice of basis functions is arbitrary, but should
be chosen to achieve good convergence with a small number of
terms.
Fergusson et al. used a set of orthogonal polynomials $q_n(k)$
to write
\begin{equation}
	S(k_1,k_2,k_3)
	=
	\sum_{p r s} \alpha_{p r s}^Q q_{(p}(k_1)q_r(k_2)q_{s)}(k_3),
	\label{eq:shape-decomposition}
\end{equation}
where bracketed indices are symmetrized with weight unity.
For details of the construction of the $q_n(k)$ we
refer to Ref.~\cite{FLS1}.

The physical region where the $k_i$ form a triangle
corresponds to a domain
$\mathcal{V}$ defined by
\begin{equation}
	2 \max ( k_1,k_2,k_3 ) \leq k_t \,,
	\label{eq:triangle-condition}
\end{equation}
where $k_t \equiv k_1 + k_2 + k_3$ is the perimeter of the
momentum triangle.
To simplify formulae it is convenient to introduce a
multi-index $n$ which runs over unique triplets $(p,r,s)$.
(By `unique', we mean triplets which generate a unique
combination $q_{(p} q_r q_{s)}$ after symmetrization.)
Defining $Q_n \equiv q_{(p}(k_1) q_r(k_2) q_{s)}(k_3)$,
we write
$S^{\text{(loc)}}_{\Phi} = \sum_n \alpha^Q_n Q_n(k_1, k_2, k_3)$.
Finally, we introduce an inner product on the physical region
by the rule
\begin{equation}
	\ipleft f, g \ipright
	\equiv
	\int_{\mathcal{V}} \d v \;
	f(k_1,k_2,k_3) g(k_1,k_2,k_3)\omega(k_1,k_2,k_3),
	\label{eq:k-ip-def}
\end{equation}
where $\d v$ is an element of area on $\mathcal{V}$
and $\omega$ is a weight function
which can be adjusted to suit our convenience.

Using this inner product we define a matrix $\gamma$ such that
\begin{align}
	\gamma_{n m}= \ipleft Q_n, Q_m \ipright.
\end{align}
The $Q_n$ are not themselves orthogonal.
Therefore,
although $\gamma_{nm}$ is symmetric,
it will not typically enjoy other special properties.
But
if the $Q_n$ have been chosen appropriately it will be
positive-definite and invertible,
in which case
the coefficients $\alpha_n^Q$ of the separable
expansion~\eqref{eq:shape-decomposition}
can be computed,
\begin{equation}
\begin{split}
	\alpha_n^Q
	& = \sum_m \ipleft S_{\Phi}^{\text{(loc)}} , Q_m \ipright
	\gamma_{m n}^{-1} .
\end{split}
\end{equation}
The accuracy of this expansion is limited by the number $\Nmax$ of
modes used.
For a shape function $S$ and an approximation $S_{N}$ using $N$ modes,
a
measure of convergence can be obtained by evaluating the ratio
\begin{equation}\label{eq:correlation}
	\mathcal{C}(S,S_{N}) \equiv
	\frac{\ipleft S, S_{N} \ipright}
	{\sqrt{\ipleft S, S \ipright \ipleft S_{N}, S_{N} \ipright}}
\end{equation}
For those models which have been studied in the literature,
only $\Or(30)$
modes are required to achieve an accuracy of at least
$90-95\%$ \cite{FLS1}.

\subsection{CMB analysis}
Given the expansion
$S_{\Phi}^{\text{(loc)}} = \sum_n \alpha_n^Q Q_n$,
the reduced CMB bispectrum~\eqref{redBisp} becomes
\begin{equation}\label{eq:primdecompcmb}
	b_{l_1 l_2 l_3}
	=
	6\fnl
	\sum_{n = (p,r,s)} \alpha_n^Q
	\int \d x \; x^2
	\tilde{q}^{(2)l_1}_{(p}(x) \tilde{q}^{(-1)l_2}_r(x) \tilde{q}^{(-1)l_3}_{s)}(x) \,,
\end{equation}
where the summation over $n$
is restricted to unique triplets,
bracketed indices are again symmetrized with
weight unity,
and the functions $\tilde{q}_p^{(2)l}$ and $\tilde{q}_p^{(-1)l}$
are defined by
\begin{equation}\label{eq:qplus2}
	\tilde{q}_p^{(2)l}(x)
	\equiv
	\frac{2}{\pi}
	\int \d k \; k^2 q_p\Big(\frac{k}{\kmax}\Big) \Delta_l(k) j_l(k x) \,,
	\quad \text{and} \quad
	\tilde{q}_p^{(-1)l}(x)
	\equiv
	\frac{2}{\pi}
	\int \d k \; k^2 P_{\Phi}(k) q_p\Big(\frac{k}{\kmax}\Big)
		\Delta_l(k) j_l(k x) \,.
\end{equation}
The separability of the expansion reduces
the integral for $b_{l_1 l_2 l_3}$ from four to two dimensions.

With current experiments, the signal-to-noise available
for each multipole is too weak to measure
the components of
$b_{l_1 l_2 l_3}$ directly.
Instead, it is conventional to construct an estimator for
the amplitude of each momentum `shape' predicted by
the underlying inflationary model.
Any such estimator sums over many components of
$b_{l_1 l_2 l_3}$ and can achieve
an acceptable signal-to-noise.
The optimal estimator
for the amplitude of a fixed bispectrum shape
$b_{l_1 l_2 l_3}$
is proportional to~\cite{Babich05}
\begin{equation}
	\mathcal{E}^{\text{opt}}
	=
	\sum_{l_i m_i} b_{l_1 l_2 l_3}
	\mathcal{G}^{l_1 l_2 l_3}_{m_1 m_2 m_3}
	\bigg[
		\Big( C^{-1} a^{\text{obs}} \Big)_{l_1 m_1}
		\Big( C^{-1} a^{\text{obs}} \Big)_{l_2 m_2}
		\Big( C^{-1} a^{\text{obs}} \Big)_{l_3 m_3}
		-
		3 C^{-1}_{l_1 m_1,l_2 m_2} a^{\text{obs}}_{l_3 m_3}
	\bigg]\,,
\end{equation}
where `obs' indicates values recovered from observation,
and $C^{-1}_{l_1 m_1,l_2 m_2}$ denotes the inverse covariance matrix.
As we have explained, it will typically be non-diagonal
because of mode-mode coupling induced by
the mask and anisotropic noise.
If we impose the diagonal approximation discussed in the
Introduction (Section~\ref{sec:introduction}),
the estimator $\mathcal{E}$ reduces to
\begin{equation}\label{eq:approxestim}
	\mathcal{E}^{\text{approx}}
	=
	\sum_{l_i m_i} b_{l_1 l_2 l_3}
	\mathcal{G}^{l_1 l_2 l_3}_{m_1 m_2 m_3}
	\frac{a_{l_1 m_1}^{\text{obs}}
		  a_{l_2 m_2}^{\text{obs}}
		  a_{l_3 m_3}^{\text{obs}}
		  -
		  3 C_{l_1 m_1,l_2 m_2} a^{\text{obs}}_{l_3 m_3}}
		  {C_{l_1}C_{l_2} C_{l_3}}\,.
\end{equation}
Its expectation value is
$
\langle \mathcal{E}^{\text{approx}}\rangle
=
\sum_{l_i}(b_{l_1 l_2 l_3}h_{l_1 l_2 l_3})^2/ (C_{l_1}C_{l_2} C_{l_3})
$.

In the next section we explain the construction of an alternative
estimator in which we sum over wavelets rather than multipoles.

\section{Review of wavelet-based estimation}\label{sec:waveletsreview}

\subsection{Definition of wavelets}

Wavelets are particularly useful
for CMB analysis due to their localization in scale and position.
In Ref.~\cite{M0111284} a continuous, isotropic
wavelet family $\psi(\vect{x}, \vect{n}, R)$
on $\mathbb{R}^2$
was constructed
from a `mother wavelet' $\Psi(\vect{x})$ by means
of translations and contractions:
$\psi(\vect{x}, \vect{n}, R) = \Psi(|\vect{x}-\vect{n}|/R)/R$.
The mother wavelet should have zero mean and decay sufficiently
fast at infinity,
\begin{align}
	\int \d \vect{x} \; \Psi(\vect{x})=0
	\qquad \text{and} \qquad
	\int \d \vect{x} \; \frac{\Psi(\vect{x})^2}{|\vect{x}|} < \infty\,.
\end{align}
Each integral is taken over $\mathbb{R}^2$.
We choose to normalize $\Psi$ so that
$\int \d\vect{x} \; \Psi(\vect{x})^2/R^2 =1$.

The wavelet transform of a function $f(\vect{x})$
with respect to location $\vect{n}$ and scale $R$ is defined by
$w(\vect{n},R) = \int f(\vect{x}) \psi(\vect{x}, \vect{n}, R)\; \d \vect{x}$.
For a sphere, the location $\vect{n} = \hat{\vect{n}}$ is a
unit vector defined by its polar and azimuthal angles.
In this paper we will exclusively use the \emph{spherical Mexican-hat
wavelet} (`SMHW'),
\begin{align}\label{eq:smhw}
	\psi_S(\theta,R) \equiv \frac{1}{\sqrt{2\pi}N(R)}\left[
		1 + \left( \frac{y}{R} \right)^2
	\right]^2
	\left[
		2 - \left( \frac{y}{R} \right)^2
	\right]
	\e{-y^2/2 R^2}\,,
\end{align}
where $N(R)=R(1+R^2/2+R^4/4)^{1/2}$ and $y=2\tan(\theta/2)$.
The SMHW depends
only on the polar angle, $\theta$, and the scale, $R$.
The Legendre transform of this wavelet, $w_l(R)$,
satisfies $\psi_S(\theta,R)=\sum_l w_l(R) P_l(\cos\theta)$~\cite{0609351}.

\subsection{Wavelets and CMB estimation}
The wavelet transform of a masked CMB temperature map
with respect to a set of $\Nscal$ scales $R_i$ is
\begin{align}\label{eq:waveletmap}
	\mathsf{W}(R_i,\hat{\bn})=\sum_{lm}a_{lm}w_l(R_i)Y_{lm}(\hat{\bn})\,,
\end{align}
where $w_l(R)$ is the Legendre transform of the SMHW.
In what follows we redefine the wavelet transform
by subtracting its mean:
we set
$W(R_i, \hat{\vect{n}}) \equiv
\mathsf{W}(R_i, \hat{\vect{n}}) -
\langle \mathsf{W}(R_i, \hat{\vect{n}}) \rangle$,
where the average is taken over
$\hat{\vect{n}}$ in the unmasked region.
This has the effect of decorrelating the $W(R_i, \hat{\vect{n}})$
on distances above the resolution scale $R_i$.
It is this property that removes the necessity to subtract a linear
correction; %for the SMHW estimator;
for further discussion see Refs.~\cite{Donzelli2012,Curto2011}.

\para{Wavelet statistics.}
The \emph{cubic wavelet statistic} is defined by
\begin{align}\label{eq:cubic}
	\tilde{W}_{ijk} & \equiv
	\frac{1}{4\pi\sigma_i \sigma_j \sigma_k}
	\int \d\hat{\vect{n}}\;
	{W}(R_i,\hat{\vect{n}})
	{W}(R_j,\hat{\vect{n}})
	{W}(R_k,\hat{\vect{n}})\,,
\end{align}
where $\sigma_i^2=\langle W(R_i,\hat{\bn})^2\rangle$,
the average again being taken over $\hat{\vect{n}}$
in the unmasked region.
For isotropic noise we have
$\sigma_i^2 = (4\pi)^{-1} \sum_{l} (2l+1)C_l w_l^2(R_i)$.
In the case of full sky coverage,
the expectation value of $\tilde{W}_{ijk}$ is given by
\begin{align}\label{eq:expecWavelet}
	V_{ijk} = \langle \tilde{W}_{ijk} \rangle =
	\frac{1}{4\pi\sigma_i \sigma_j \sigma_k}
	\sum_{l_1, l_2, l_3} w_{l_1}(R_i) w_{l_2}(R_j) w_{l_3}(R_k)
	h_{l_1 l_2 l_3}^2 b_{l_1 l_2 l_3}\, ,
\end{align}
where $b_{l_1 l_2 l_3}$ is the reduced bispectrum~\eqref{redBisp}
corresponding to the primordial shape
whose amplitude we wish to constrain;
see Eqs.~\eqref{bispLs} and~\eqref{eq:reduced-bispectrum}.
It would ordinarily be computed from the bispectrum
predicted by a microscopic inflationary model
as described in Section~\ref{sec:primtolate}.
The quantity
$h_{l_1 l_2 l_3}^2$ satisfies
\begin{align}\label{eq:hstuff}
	h_{l_1 l_2 l_3}^2 \equiv \frac{(2l_1+1)(2l_2+1)(2l_3+1)}{8\pi}
	\int_{-1}^1 \d\mu \; P_{l_1}(\mu)P_{l_2}(\mu)P_{l_3}(\mu)\,.
\end{align}
Therefore,
given the wavelet transform $W(R_i, \hat{\vect{n}})$,
the computation of $V_{ijk}$ for a generic %non-Gaussian
bispectrum will involve
$\Or(\Nscal \lmax^3)$ operations.
For a real experiment some of the sky must be masked
and~\eqref{eq:expecWavelet} no longer applies.
In this case
the expectation value of the cubic statistic for each scale
must be found using simulations.
We describe how suitable
simulations incorporating the underlying bispectrum can be performed
in Section~\ref{sec:primtolate}.

\para{Optimal estimator.}
The optimal estimator for the amplitude
of the bispectrum shape of interest, $b_{l_1 l_2 l_3}$, is%
	\footnote{In writing this formula we have assumed
	that the bispectrum $b_{l_1 l_2 l_3}$ used
	to compute $V_{ijk}$ is normalized so that $\fnl^b = 1$.}
\cite{Curto0807}
\begin{align}\label{eq:fnlWavelet}
	\fnlhat^b \equiv
	\frac{\sum_{ijkrst} {V}_{ijk}C^{-1}_{ijk,rst} \tilde{W}_{rst}}
	{\sum_{ijkrst} {V}_{ijk}C^{-1}_{ijk,rst} {V}_{rst}}\,,
\end{align}
where $C_{ijk,rst}$ is the covariance matrix of the cubic
statistics $\tilde{W}_{ijk}$.
In terms of the estimator $\mathcal{E}$
introduced in Section~\ref{sec:modesreview}
it has the schematic form $\fnlhat = \mathcal{E} / \langle \mathcal{E} \rangle$.
We write $\fnlhat^b$ to indicate that this amplitude depends on
the bispectrum shape used to construct the estimator.
It does not coincide with the traditional definition of
Spergel \& Komatsu~\cite{KomatsuSpergel2001}
unless the bispectrum $b_{l_1 l_2 l_3}$
corresponds to the local model.

In~\eqref{eq:fnlWavelet} it is not necessary to include
the linear term
\begin{align}\label{eq:waveletlinear}
	\tilde{W}_{ijk}^{\text{linear}} =
	\frac{1}{4\pi\sigma_i \sigma_j \sigma_k}
	\int \d\hat{\vect{n}} \; {W}(R_i,\hat{\bn})
	\langle{W}(R_j,\hat{\bn}){W}(R_k,\hat{\bn})\rangle
	+ \text{2 perms}\,,
\end{align} 
because of the scale-by-scale subtraction
of the mean
for each
wavelet coefficient~\cite{Curto2011}.
Nevertheless,
for completeness,
we will also subtract this term unless
otherwise stated and understand
$\tilde{W}_{ijk}\rightarrow \tilde{W}_{ijk}-\tilde{W}_{ijk}^{\text{linear}}$.
Eq.~\eqref{eq:fnlWavelet} can be written more succinctly as
\begin{align}\label{eq:fnlWavelet2}
	\fnlhat = \frac{\sum_{I J} {V}_{I}C^{-1}_{IJ} \tilde{W}_{J}}
	{\sum_{KL} {V}_{K}C^{-1}_{KL} {V}_{L}}\, ,
\end{align}
where $I, J, K, L$ are understood as multi-indices,
each ranging over a three-component tuple.
Therefore $I = (i,j,k)$, and similarly for $J$, $K$, $L$.
The Fisher estimate for
the variance of $\fnlhat$ is
\begin{align}\label{eq:varfnl}
	\sigma_F^2(\fnlhat) \equiv
	\frac{1}{\sum_{KL} {V}_{K}C^{-1}_{KL} {V}_{L}}\,.
\end{align}

To keep numerical errors in the inverse covariance matrix under
control we carry out the calculation using principal component analysis.
Due to the vastly reduced dimensionality---%
we use of order $10^3$ cubic statistics---%
the computation
is much faster
than that of the pixel-by-pixel inverse covariance matrix needed
for optimization of bispectrum-based estimators.

\para{Normalized amplitude estimate.}
Instead of Eq.~\eqref{eq:fnlWavelet} we can consider an alternative measure of the
amplitude, normalized to the local shape.
This redefined measure was introduced in Ref.~\cite{FLS1}, where it was denoted $\Fnl$.
The estimator is
\begin{equation}
	\Fnlhat^b =
	\frac{\sum_{IJ} V_I C^{-1}_{IJ} \tilde{W}_J}
	{\left( \sum_{KL} V_K C^{-1}_{KL} V_L \right)^{1/2}
	 \left( \sum_{KL} V_M^{\text{loc}} C^{-1}_{MN} V_N^{\text{loc}} \right)^{1/2}} ,
	\label{eq:capital-fnl-wavelet}
\end{equation}
where the $V_I^{\text{loc}}$ should be computed using the local bispectrum shape.
The estimators~\eqref{eq:fnlWavelet} and~\eqref{eq:capital-fnl-wavelet}
contain identical information and differ only in their normalization.
Eq.~\eqref{eq:capital-fnl-wavelet} is less useful for comparison to specific models
because it does not yield a correctly-normalized estimate of each amplitude
appearing in the primordial bispectrum~\eqref{eq:b-def}.

\section{Projection from primordial to CMB bispectra and map-making}\label{sec:primtolate}

\subsection{From primordial to CMB bispectra}

The CMB bispectrum $b_{l_1 l_2 l_3}$ [sometimes called the `late-time' bispectrum
to distinguish it from the primordial bispectrum of Eq.~\eqref{eq:b-def}]
is a function of the multipoles $l_1$, $l_2$ and $l_3$ and therefore may \emph{also}
be decomposed in terms of the partial-wave basis $Q_n$.
To do so, we define a weighted copy of the bispectrum $s(l_1, l_2, l_3)$ and introduced
`barred' coefficients $\baralpha_n^Q$ such that
\begin{equation}\label{eq:latetime}
	s_{l_1 l_2 l_3}
	\equiv
	\frac{(2l_1+1)^{1/6}(2l_2+1)^{1/6}(2l_3+1)^{1/6}}{\sqrt{C_{l_1}C_{l_2}C_{l_3}}}
	b_{l_1l_2 l_3}
	\equiv
	\sum_n \baralpha_n^Q Q_n(l_1,l_2,l_3)\,.
\end{equation}
The choice of weighting was explained in Ref.~\cite{FLS1}.
The barred (`late-time') coefficients should be carefully distinguished from the
unbarred (`primordial') coefficients which appear in Eq.~\eqref{eq:shape-decomposition}.
For notational simplicity it is sometimes helpful to renormalize the partial-wave
basis by introducing new functions $b_{l_1 l_2 l_3}^{(n)}$ which satisfy
\begin{equation}
	b_{l_1 l_2 l_3}^{(n)}
	\equiv
	\frac{\sqrt{C_{l_1}C_{l_2} C_{l_3}}}
		 {(2 l_1+1)^{1/6} (2 l_2+1)^{1/6} (2 l_3+1)^{1/6}}
	Q_n(l_1, l_2, l_3) .
	\label{eq:bln}
\end{equation}
In terms of this basis we have
$b_{l_1 l_2 l_3} = \sum_n \baralpha_n^Q b_{l_1 l_2 l_3}^{(n)}$.

\para{Primordial to late-time mapping.}
We now aim to express the late-time coefficients $\baralpha^Q_n$ in terms of their
primordial counterparts.
To do so we can project $s_{l_1 l_2 l_3}$ on to successive basis functions
$Q_n(l_1, l_2, l_3)$. However, the inner product~\eqref{eq:k-ip-def}
is no longer appropriate because the multipole labels $l_i$ are discrete.
Therefore we introduce a new inner product (see Ref.~\cite{FLS1}) defined by 
\begin{equation}
\begin{split}
	\ipleft f, g \ipright
	& \equiv
	\sum_{l_1 l_2 l_3} f(l_1,l_2,l_3) g(l_1,l_2,l_3)
	\frac{h_{l_1 l_2 l_3}^2}{(2l_1+1)^{1/3}(2l_2+1)^{1/3}(2l_3+1)^{1/3}}
	\\
	& =
	\frac{1}{8\pi}
	\sum_{l_1 l_2 l_3}
	\int_{-1}^1 \d\mu \;
	f(l_1,l_2,l_3) g(l_1,l_2,l_3)
	(2l_1+1)^{2/3}(2l_2+1)^{2/3}(2l_3+1)^{2/3}
	P_{l_1}(\mu)P_{l_2}(\mu)P_{l_3}(\mu)\,.
\end{split}
\label{eq:l-ip-def}
\end{equation}

We first use~\eqref{eq:primdecompcmb} to express $b_{l_1 l_2 l_3}$,
and hence $s_{l_1 l_2 l_3}$,
in terms of the
primordial coefficients $\alpha^Q_n$.
Projecting the resulting expression onto $Q_m$ gives
\begin{equation}\label{eq:decomp1}
	\ipleft s_{l_1 l_2 l_3}, Q_m(l_1,l_2,l_3) \ipright
	=
	\sum_n \alpha_n^Q
	\int \d x \; x^2 \tilde{\gamma}_{n m}(x)\,,
\end{equation}
where
\begin{equation}
	\tilde{\gamma}_{n m}(x)
	\equiv \frac{1}{8\pi} \int_{-1}^1 \d\mu \;
	N^{(2)}_{(n_1 m_1}(\mu,x) N^{(-1)}_{n_2 m_2}(\mu,x) N^{(-1)}_{n_3 m_3)}(\mu,x)\,.
\end{equation}
In this expression the multi-index $n$ is the triple $(n_1, n_2, n_3)$;
the multi-index $m$ is the triple $(m_1, m_2, m_3)$;
the bracketed indices $( \cdots )$ denote simultaneous symmetrization
over both triples with weight unity;
and we have defined
\begin{equation}\label{eq:gammatilde}
\begin{split}
	N^{(2)}_{n_1 m_1}(\mu,x)
	& \equiv
	\sum_{l} \frac{(2 l+1)^{5/6}}{\sqrt{C_{l}}}
	\tilde{q}_{n_1}^{(2)l}(x) q_{m_1}\Big(\frac{l}{\lmax}\Big)P_{l}(\mu)\,,
	\\
	N^{(-1)}_{n_1 m_1}(\mu,x)
	& \equiv
	\sum_{l} \frac{(2 l+1)^{5/6}}{\sqrt{C_{l}}}
	\tilde{q}_{n_1}^{(-1)l}(x)q _{m_1}\Big(\frac{l}{\lmax}\Big)P_{l}(\mu)\,.
\end{split}
\end{equation}

Alternatively, projecting the late-time decomposition on the right-hand side
of Eq. \eqref{eq:latetime} gives
\begin{equation}\label{eq:decomp2}
	\ipleft s_{l_1 l_2 l_3}, Q_m(l_1,l_2,l_3) \ipright
	=
	\frac{1}{8\pi}
	\sum_n \baralpha_n^Q
	\int_{-1}^1 \d\mu \;
	\barN_{(n_1 m_1}(\mu) \barN_{n_2 m_2}(\mu) \barN_{n_3 m_3)}(\mu)
	=
	\sum_n \baralpha_n^Q \bargamma_{n m}\,,
\end{equation}
where the same conventions apply for bracketed indices, and
\begin{equation}\label{eq:gammabar}
	\barN_{n_1 m_1}(\mu)
	=
	\sum_{l}{(2 l+1)^{2/3}}
		{q}_{n_1}\Big(\frac{l}{\lmax}\Big)
		q_{m_1}\Big(\frac{l}{\lmax}\Big)
		P_{l}(\mu)\,,
	\quad \text{and} \quad
	\bargamma_{n m}
	=
	\frac{1}{8\pi}
	\int_{-1}^1 \d\mu \;
	\barN_{(n_1 m_1}(\mu) \barN_{n_2 m_2}(\mu) \barN_{n_3 m_3)}(\mu)\,.
\end{equation}
Equating Eqs.~\eqref{eq:decomp1} and~\eqref{eq:decomp2}
gives the required relationship between primordial and late-time coefficients
\begin{equation}\label{eq:Gammapn}
	\baralpha_n^Q
	=
	\sum_{p m} \alpha_p^Q
	\Big(
		\int \d x \; x^2 \tilde{\gamma}_{p m}(x)
	\Big)
	\bargamma^{-1}_{m n}
	\equiv
	\sum_p \alpha_p^Q \Gamma_{p n}\,,
\end{equation}
and $\Gamma_{p n}$ is defined by this expression.
It can be regarded as a projection of the transfer function for the bispectrum into
mode-space, and describes the change in shape from
the primordial era
(given by the coefficients $\alpha_n^Q$)
to the surface of last-scattering (given by the $\baralpha_n^Q$).

The $Q_n$ were not specifically constructed to give a good representation
of the \emph{angular} bispectrum,
and therefore
one might harbour some reservations that the
approximation of $b_{l_1 l_2 l_3}$ by the same
number of basis functions used to represent the
\emph{primordial} bispectrum may introduce
an unwanted error.
However, in practice Eq.~\eqref{eq:Gammapn}
proves to be extremely accurate, typically producing better than $99\%$ correlation
with $\Or(100)$ modes.
For further details and discussion, see Ref.~\cite{FRS2}.

\subsection{Simulating non-Gaussian maps}\label{subsec:sims}
It was explained in Section~\ref{sec:waveletsreview}
that, for a real experiment,
the effects of sky masking and anisotropic noise mean that
the expectation values $V_I = \langle \tilde{W}_I \rangle$ for each cubic wavelet
statistic must be obtained by numerical simulation.
[See discussion under Eq.~\eqref{eq:hstuff}.]
In Ref.~\cite{FLS1}, a simple prescription was given to carry out such
simulations.
We set $a_{lm}=a^G_{lm}+ F_{\rm{NL}} a^B_{lm}$, where
$a^G_{lm}$ is the Gaussian part of each CMB multipole
and $a^B_{lm}$ is a non-Gaussian correction,
\begin{equation}
	a^B_{lm}
	\equiv \frac{1}{6}
	\sum_{l_2 m_2}\sum_{l_3 m_3} b_{l l_2 l_3}
	\mathcal{G}^{l l_2 l_3}_{m m_2 m_3}
	\frac{a^{G*}_{l_2 m_2}a^{G*}_{l_3 m_3} }{C_{l_2} C_{l_3}}\,.
\end{equation}
It follows that
\begin{equation} 
	a^B_{lm}
	=
	\sum_n \baralpha^Q_n a^{B (n)}_{lm}
	=
	\sum_n \frac{\baralpha^Q_n}{6} \sum_{l_2 m_2} \sum_{l_3 m_3}
	b^{(n)}_{l l_2 l_3} \mathcal{G}^{l l_2 l_3}_{m m_2 m_3}
	\frac{a^{G*}_{l_2 m_2}a^{G*}_{l_3 m_3} }{C_{l_2} C_{l_3}}\,.
\end{equation}
The $a^{B (n)}_{lm}$ can be computed very efficiently, since
% efficient because does not require a line of sight integral
% - find a way to explain this
\begin{equation}
	a^{B (n)}_{lm}
	=
	\frac{1}{6}
	\frac{\sqrt{C_l}}{(2 l+1)^{1/6}}
	\int \d\hat{\vect{n}} \;
	Y_{ lm}(\hat{\vect{n}})
	q_{(p}(l/\lmax) M_r^G(\hat{\vect{n}}) M_{s)}^G(\hat{\vect{n}})\,,
\end{equation}
where the multi-index $n$ is the triple $(p,r,s)$ and
the weighted maps $M_p^G(\hat{\bn})$ are defined as
\begin{equation}
	M_p^G(\hat{\vect{n}})
	=
	\sum_{l m} q_p\Big(\frac{l}{\lmax}\Big) \frac{a^G_{lm}}{(2 l+1)^{1/6}\sqrt{C_l}} Y_{lm}(\hat{\vect{n}})\,.
\end{equation}

\section{Application of the modal approach to wavelets}\label{sec:waveletsmodes}

We are now in a position to connect the modal and wavelet
approaches. In particular, we wish to use the modal decomposition of
some specific primordial bispectrum shape $S^{\text{(loc)}}_{\Phi}$
(specified by its coefficients $\alpha^Q_n$)
to determine the expectation value $V_I$ for each cubic wavelet statistic.
Once these expectation values are determined,
the formalism of wavelet estimators described in
Section~\ref{sec:waveletsreview} can be used to recover the amplitude
with which $S^{\text{(loc)}}_{\Phi}$ appears in the data with a near-optimal
error bar.

To do so, we write $V_I =\sum_n \baralpha_n^Q V_{n I}$.
The matrix $V_{n I}$ can be thought of as a change of basis from partial-waves to
wavelets and must be computed using the prescription given in Section~\ref{subsec:sims}
for evaluation of a non-Gaussian map.
We find%
	\footnote{To simplify notation we have omitted
	the dependence of each wavelet map on $\hat{\vect{n}}$.}
\begin{equation}\label{eq:vni}
	V_{n I}
	=
	\frac{1}{4 \pi \sigma_i \sigma_j \sigma_k}
	\int \d\hat{\vect{n}} \;
	\langle
		W^G(R_i) W^G(R_j) W^{B(n)}(R_k)
		+ W^G(R_i) W^{B(n)}(R_j) W^G(R_k)
		+ W^{B(n)}(R_i) W^G(R_j) W^{G}(R_k)
	\rangle\,,
\end{equation}
where the wavelet maps $W^G$ and $W^{B(n)}$ are
given by \eqref{eq:waveletmap}, with $a_{lm}$ replaced by $a_{lm}^G$ and $a_{lm}^{B(n)}$, respectively. The scale-by-scale mean should be subtracted out as usual.
Note that the $V_{n I}$ are independent of any model-specific details,
such as the shape of the bispectrum for which we are trying to construct
an estimator, and can be precomputed.

\para{Wavelet estimator.}
It is now possible to write down
the wavelet estimator for the amplitude of a bispectrum shape specified by
primordial coefficients $\alpha^Q_n$ and late-time coefficients $\baralpha^Q_n$.
It is
\begin{equation}\label{eq:WaveletAlphas}
	\fnlhat^b =
	\frac{\sum_n \baralpha_n^Q \sum_{IJ} V_{n I} C^{-1}_{IJ}\tilde{W}_J}
		 {\sum_{n m} \baralpha_n^Q  \baralpha_m^Q \sum_{IJ} V_{n I} C^{-1}_{IJ}V_{m J} }\,.
\end{equation}
If desired, this can be written in a form similar to Ref.~\cite{FLS1}
\begin{equation}\label{eq:fnlhatother}
	\fnlhat^b =
	\frac{\sum_n \baralpha_n^Q \barbeta^Q_n}
		 {\sum_{n m} \baralpha_n^Q \baralpha_m^Q \bargamma_{nm}}\,,
\end{equation}
where we have defined
$\barbeta^Q_n = \sum_{IJ} V_{nI} C^{-1}_{IJ} \tilde{W}_J$ and
$\bargamma_{nm} = \sum_{IJ} V_{nI} C^{-1}_{IJ} V_{m J}$.

For any particular experiment, the quantities $\barbeta^Q_n$ and $\bargamma_{mn}$
are model-independent and can be precomputed. (We emphasize that they vary between
experiments due to the details of noise and masking.) Once these
coefficients are available, the estimator $\fnlhat^b$
for any primordial model can be obtained by trivial summations.

\para{Orthogonalized modes.}
Although~\eqref{eq:WaveletAlphas} is our final result for the wavelet estimator,
it can be rewritten in an equivalent form which orthogonalizes the partial-wave basis.
We perform a Cholesky decomposition of the matrix
$\bargamma_{nm}$ to obtain
$\bargamma_{nm} = \sum_r \lambda^{-1}_{n r}\lambda^{-1}_{m r}$.
Defining
$\baralpha_r^R \equiv \sum_n \lambda^{-1}_{n r}\alpha_n^Q$
and $\barbeta^R_r \equiv \sum_r \lambda_{r n}\barbeta^Q_r$, it follows that
\begin{equation}\label{eq:fnlconst}
	\fnlhat^b =
	\frac{\sum_n \baralpha_n^R \barbeta^R_n }{\sum_{n} (\baralpha_n^R)^2}\,.
\end{equation}
This expression is particularly useful because it allows us to deduce that the expectation
values $\langle \barbeta^R_n \rangle$ obtained from an ensemble of maps with
bispectrum $b_{l_1 l_2 l_3}$
satisfy the relation
\begin{equation}\label{eq:sanity}
	\langle \barbeta_n^R \rangle = \baralpha_n^R\,.
\end{equation}
This relation can be used as a `sanity check' for simulations of a specific model.

Alternatively, the analysis could be carried out
entirely in wavelet space. Introducing a Cholesky decomposition
of the inverse covariance matrix
$C^{-1}_{IJ}=\sum_K L_{IK}L_{JK}$
we may similarly define
$\fnlhat^b =\sum_K A_K B_K/\sum_K A_K^2$,
where $A_K=\sum_I L_{IK}V_I$ and $B_K=\sum_J L_{JK}\tilde{W}_J$.
The `sanity-check' given by Eq.~\eqref{eq:sanity}
now becomes $\langle B_K\rangle = A_K$.

\section{WMAP7 implementation}\label{sec:wmaptest}

In Sections~\ref{sec:modesreview}--\ref{sec:waveletsmodes} we have
assembled the theoretical framework needed to construct wavelet estimators
for any chosen primordial bispectrum.
In this Section and the next
we apply this formalism to the coadded $V + W$ foreground-cleaned
maps from the WMAP 7-year data release, working up to $\lmax = 1000$.
The data is at a resolution of $6.9$ arcmin, corresponding to
$N_{\text{side}}=512$ for
HEALPix \cite{0409513}.

Previous analyses of this dataset have used $\lmax = 1500$.
However, the purpose of this paper is to provide a proof-of-concept for the
combined modal/wavelet methodology, rather than to obtain the most stringent
possible error bar.
In any case, the bispectrum analysis in Ref.~\cite{FLS10}, which employed
modal techniques, was carried only to $\lmax = 500$.
The authors of that paper noted that the pseudo-optimal approach of Ref.~\cite{FS3}
tends to saturate for larger $l$.

For completeness,
in Table~\ref{table:parameters} we list the cosmological parameters used in this analysis.
The primordial power spectrum is parametrized as a power-law with
$P_{\Phi}(k) = A_{\Phi} k^{-3} ( k / k_\star )^{n_s - 1}$, and the pivot
scale $k_\star$ is taken to be $k_\star = 0.002h \; \text{Mpc}^{-1}$.
\begin{table}[htp]
\begin{center}
	\heavyrulewidth=.08em
	\lightrulewidth=.05em
	\cmidrulewidth=.03em
	\belowrulesep=.65ex
	\belowbottomsep=0pt
	\aboverulesep=.4ex
	\abovetopsep=0pt
	\cmidrulesep=\doublerulesep
	\cmidrulekern=.5em
	\defaultaddspace=.5em
	\renewcommand{\arraystretch}{1.6}

	\begin{tabular}{SS}
		
		\toprule
	
		\multicolumn{1}{c}{parameter} & \multicolumn{1}{c}{value} \\
		\Omega_b h^2 & 0.0227 \\
		\Omega_c h^2 & 0.1116 \\
		\Omega_{\Lambda} & 0.729 \\
		\tau & 0.085 \\
		A_{\Phi} & 1.736 \times 10^{-8} \\
		n_s & 0.963 \\
		
		\bottomrule
	\end{tabular}
\end{center}
\caption{Cosmological parameters used in the WMAP7 analysis.\label{table:parameters}}
\end{table}

\subsection{Generating the wavelets and corresponding masks}\label{subsec:waveletsgen}

The spherical Mexican-hat wavelet is defined in Fourier space by
the Legendre transform of Eq.~\eqref{eq:smhw}.
For each map, the wavelet coefficients are obtained by the convolution
in Eq.~\eqref{eq:waveletmap}.
We use the same fifteen scales chosen by Curto et al.~\cite{Curto2011},
and extend the WMAP KQ75 mask appropriately for each scale.
We list the angular scales in Table~\ref{tab:maskcoverage} together with the
fraction of sky available at each scale after applying the mask.

\begin{figure}
\centering
\begin{tabular}{c c c}
$R_0$ (unconvolved) & $R_1=2.9'$ & $R_2=4.5'$\\
\includegraphics[width=0.3\linewidth] {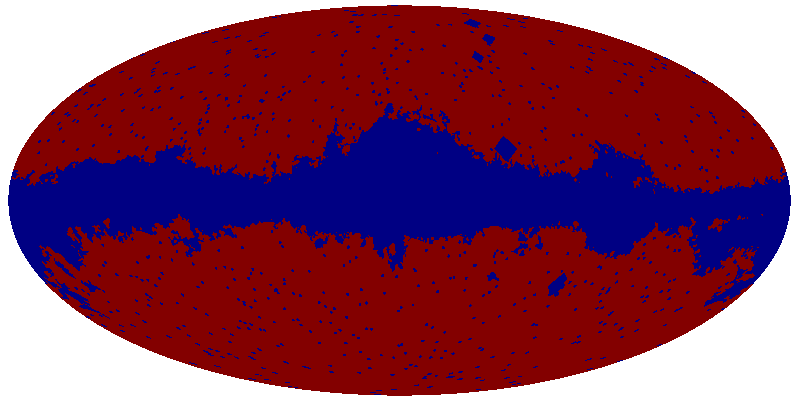}
 & 
 \includegraphics[width=0.3\linewidth] {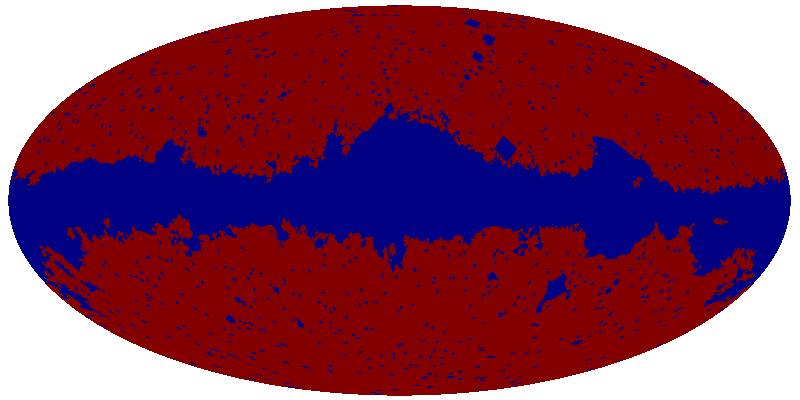}
 & 
 \includegraphics[width=0.3\linewidth] {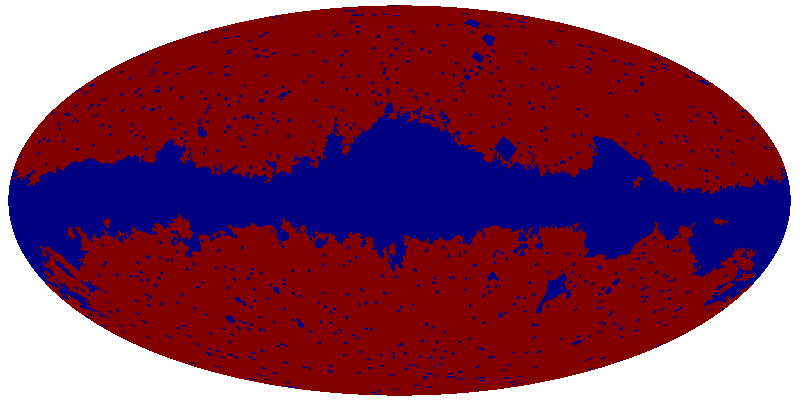}\\
 $R_3=6.9'$ & $R_4=10.6'$ & $R_5=16.3'$\\
 \includegraphics[width=0.3\linewidth] {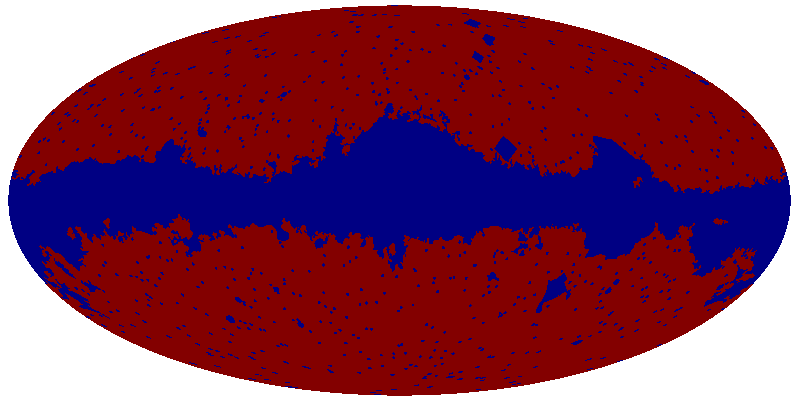}
 & 
 \includegraphics[width=0.3\linewidth] {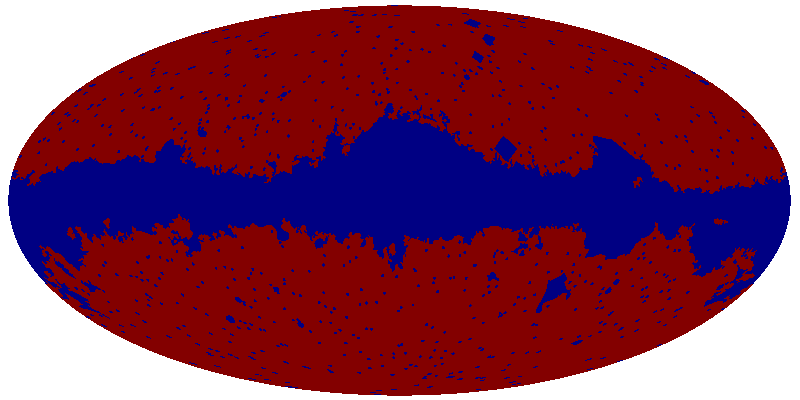}
 & 
 \includegraphics[width=0.3\linewidth] {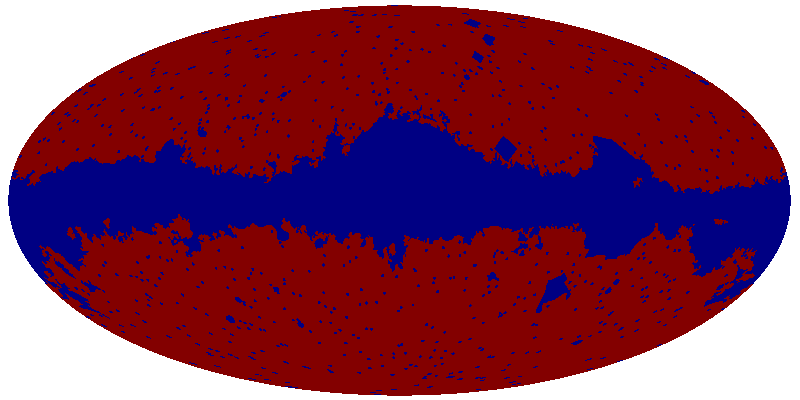}\\
 $R_6=14.9'$ & $R_7=38.3'$ & $R_8=58.7'$\\
 \includegraphics[width=0.3\linewidth] {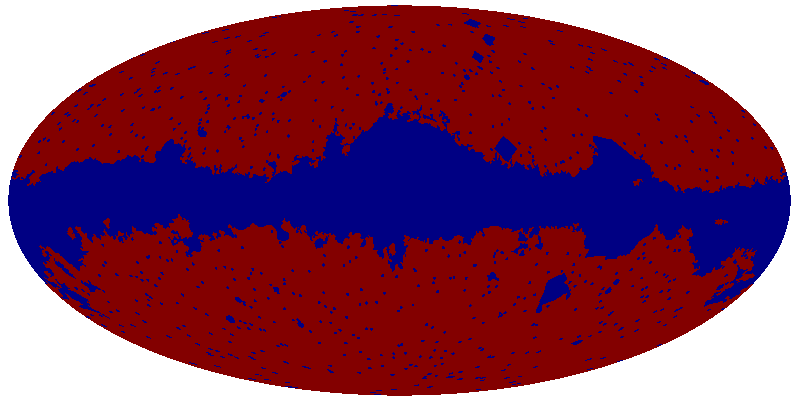}
 & 
 \includegraphics[width=0.3\linewidth] {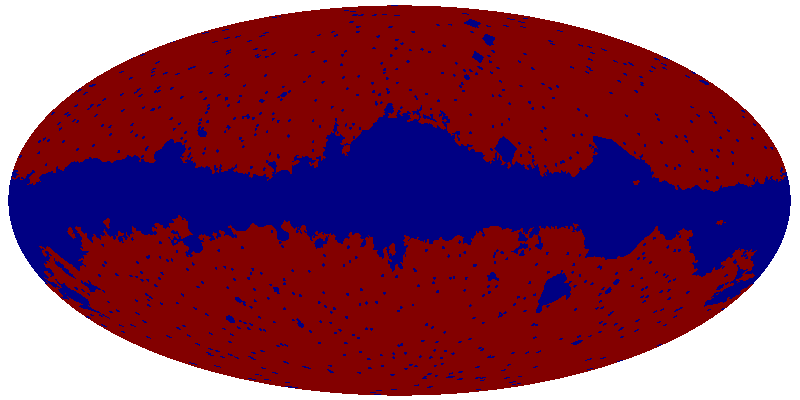}
 & 
 \includegraphics[width=0.3\linewidth] {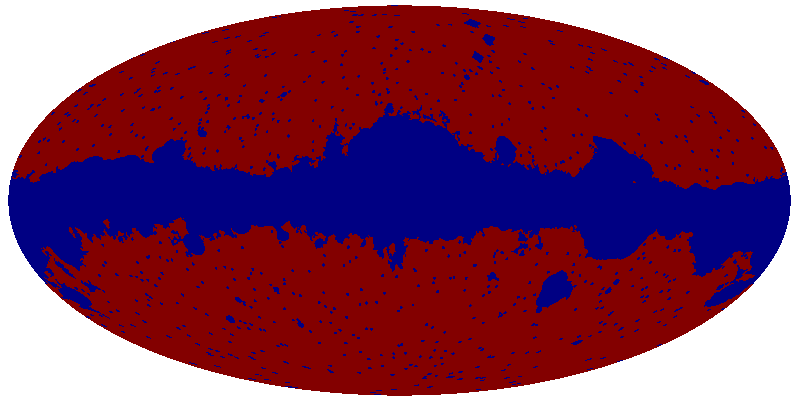}\\
 $R_9=90.1'$ & $R_{10}=138.3'$ & $R_{11}=212.3'$\\
 \includegraphics[width=0.3\linewidth] {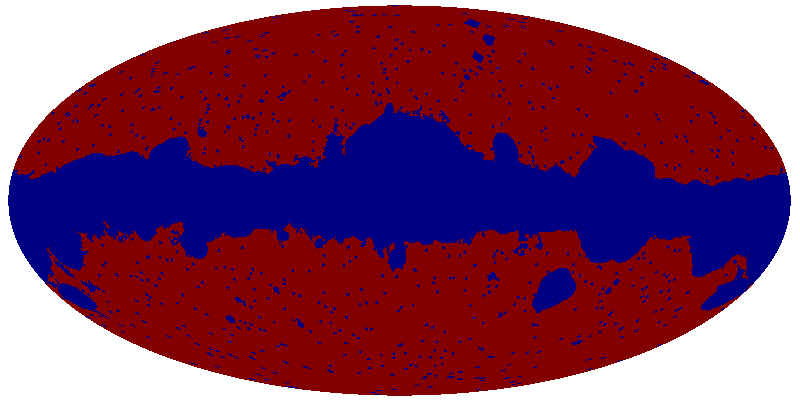}
 & 
 \includegraphics[width=0.3\linewidth] {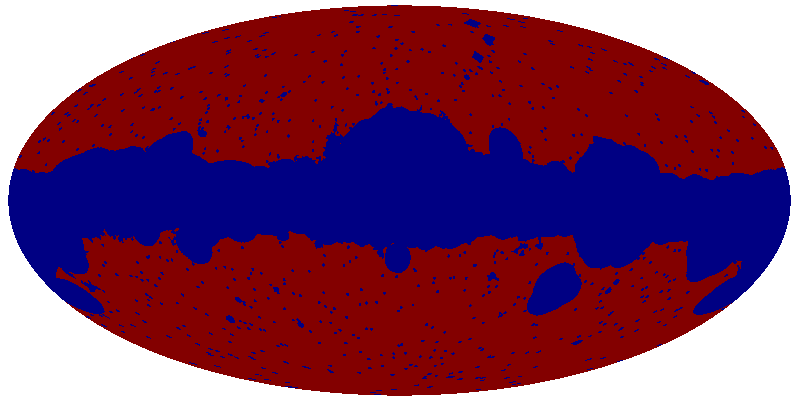}
 & 
 \includegraphics[width=0.3\linewidth] {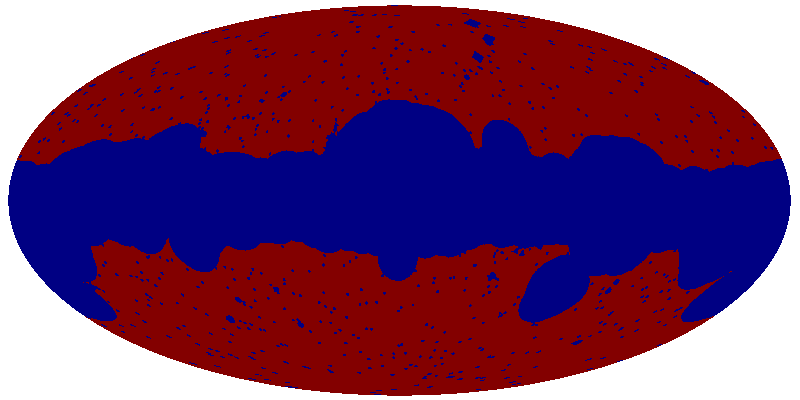}\\
 $R_{12}=325.8'$ & $R_{13}=500'$ & $R_{14}=767.3'$\\
 \includegraphics[width=0.3\linewidth] {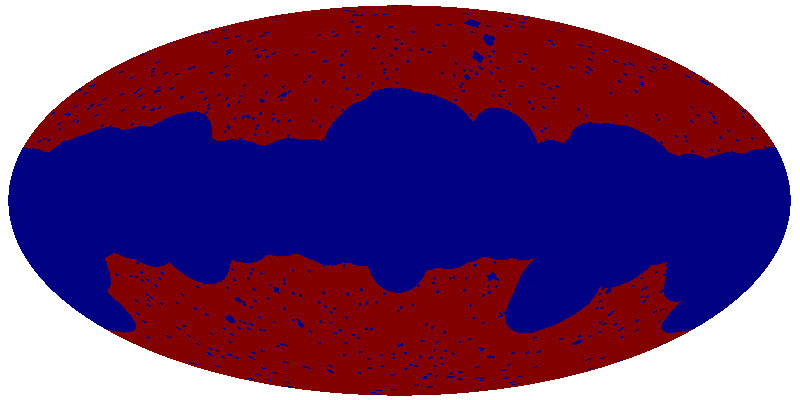}
 & 
 \includegraphics[width=0.3\linewidth] {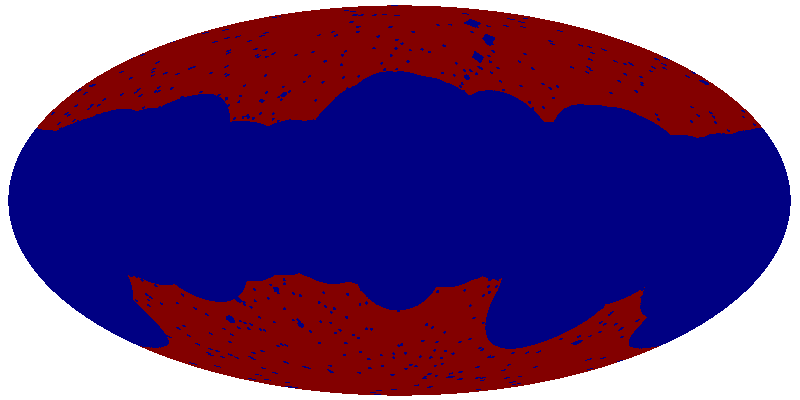}
 & 
 \includegraphics[width=0.3\linewidth] {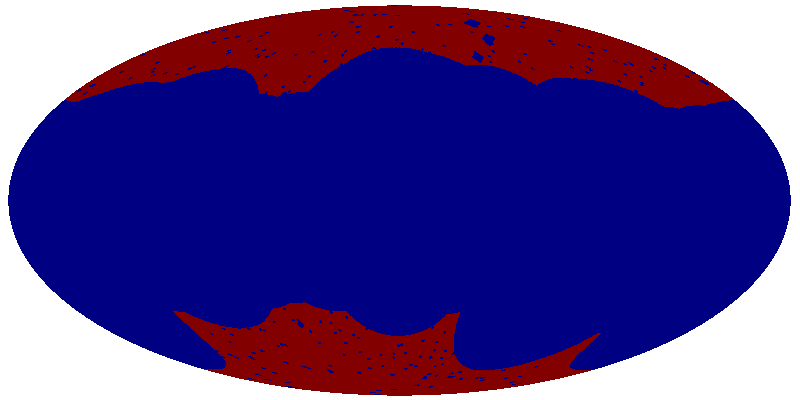}
\end{tabular}
\caption{(Extended) masks appropriate for the analysis at each wavelet scale.
The proportion of sky coverage for each mask is summarized in
Table~\ref{tab:maskcoverage}.}
\label{fig:masks}
\end{figure}

The construction
of an appropriate mask
can be done in various ways. Here, our results correspond to masks constructed by
taking the KQ75 mask without point sources and extending it so that,
for the wavelet of scale $R$, any pixel within
$2.5R$ of a masked pixel is excluded from the analysis.
For small-scale wavelets (up to $R_6$) we superpose the mask around point sources.
On small scales this mask is believed to be sufficiently extended not to cause
contamination in the wavelet coefficients. On large scales the effect is negligible.
Therefore, further extension of the mask for any $R_i$ would be too
conservative.
We plot the extended masks in Figure~\ref{fig:masks}.

Constructing extended masks by convolving the existing mask with the wavelet at
each scale and leaving out regions where the coefficients are contaminated by more
than $1\%$ results in very similar masks.

\begin{table}[ht]
	\heavyrulewidth=.08em
	\lightrulewidth=.05em
	\cmidrulewidth=.03em
	\belowrulesep=.65ex
	\belowbottomsep=0pt
	\aboverulesep=.4ex
	\abovetopsep=0pt
	\cmidrulesep=\doublerulesep
	\cmidrulekern=.5em
	\defaultaddspace=.5em
	\renewcommand{\arraystretch}{1.6}
\begin{center}
\begin{tabular}{csSsSsSsSsSsSsSs}

  \toprule

  Wavelet scale &
  R_0 &
  R_1 &
  R_2 & 
  R_3 &
  R_4 & 
  R_5 &
  R_6 & 
  R_7 &
  R_8 & 
  R_9 & 
  R_{10} & 
  R_{11} & 
  R_{12} & 
  R_{13} &
  R_{14} \\
  Angular scale &
  0 &
  2.9' &
  4.5' &
  6.9' &
  10.6' &
  16.3' &
  24.9' &
  38.3' &
  58.7' &
  90.1' &
  138.3' &
  212.3' &
  325.8' &
  500' &
  767.3' \\
  Sky coverage ($\%$) &
  70.6 &
  70.6 &
  70.6 &
  70.6 &
  70.6 &
  70.5 &
  70.4 &
  70.1 &
  69.3 &
  67.3 &
  63.5 &
  57.3 &
  48.4 &
  36.2 &
  20.6 \\

  \bottomrule
\end{tabular}
\end{center}
\caption{Proportion of sky covered at each wavelet scale $R_0$ to $R_{14}$.
%The corresponding scales are summarised in Section~\ref{subsec:waveletsgen}.
The mask at scale $R_0$ corresponds to the KQ75 mask.}
\label{tab:maskcoverage}
\end{table}

\subsection{Calculating the modal coefficients}

As we have explained in Section~\ref{subsec:primdecomp},
to compute the late-time coefficients ($\baralpha^Q_n$)
we first compute the primordial coefficients ($\alpha_n^Q$)
for the shape $S_{\Phi}^{\text{(loc)}}$ defined in~\eqref{eq:shapeloc}.
For each triple $(p, r, s)$ which defines a basis function we compute
$\ell^2 = p^2 + r^2 + s^2$ and retain only the 80 modes with lowest $\ell$.
This is sufficient to obtain a correlation of $\gtrsim 99\%$ for the usual
bispectrum templates (local, constant, equilateral, orthogonal and flattened),
except for the flattened model where we achieve $\sim 95\%$
correlation. This can be attributed to an inherent lack of smoothness
of the template in the flattened limit, which makes the modal decomposition
converge only slowly.
We use the same number of modes to decompose the reduced angular bispectrum
$b_{l_1 l_2 l_3}$.
Later, we will verify that this is sufficient to ensure an accurate
representation.

In order to calculate the transfer matrix $\Gamma_{pn}$
from primordial to late-time coefficients
we first extract the transfer function from CAMB~\cite{9911177}
and
compute the line-of-sight projections
$\tilde{q}_p^{(2) l}(x)$ and $\tilde{q}_p^{(-1) l}(x)$ defined
in Eq.~\eqref{eq:qplus2}.
We then compute $\tilde{\gamma}_{mn}(x)$ and $\bargamma_{mn}$
using Eqs.~\eqref{eq:gammatilde} and~\eqref{eq:gammabar}.
Combining all these elements
enables us to compute the transfer matrix from~\eqref{eq:Gammapn}.

\subsection{Calculating the observed wavelets and modes}

Our first task is to estimate the expectation value $V_I$ of each cubic wavelet
statistic.
To do so
we simulate Gaussian spherical harmonic transforms
with the variance for each multipole given by the angular temperature power
spectrum, $C_l$. These are denoted $a_{lm}^G$.
Using the prescription outlined in Section~\ref{subsec:sims} we
generate the simulated non-Gaussian multipoles, $a_{lm}^{B(n)}$,
corresponding to the bispectrum
basis function $b_{l_1 l_2 l_3}^{(n)}$ [defined in~\eqref{eq:bln}].
The combination $a_{lm} = a_{lm}^G + a_{lm}^{B(n)}$ gives the simulated temperature
map.
The WMAP7 beam, $b_l$, and noise $n_{l m}$ can be incorporated via the transformation
\begin{equation}
	a_{lm} \rightarrow \tilde{a}_{lm} = b_l a_{lm} + n_{l m} .
\end{equation}
Using Eq.~\eqref{eq:waveletmap} we create simulated Gaussian and non-Gaussian
wavelet maps for each scale and apply the appropriate masks.
Then, for each map, the average in the unmasked region is subtracted.
Finally we extract the cubic statistics $W^G_I$, $W^{B(n)}_I$
using~\eqref{eq:cubic}.
(Note here that the normalization coefficients for
each cubic statistic, $\sigma_i^2$, are obtained using the assumption of
isotropic noise. One should not be concerned, because these factors
merely represent a normalization convention and
cancel out in the estimator, Eq.~\eqref{eq:fnlWavelet2}.)
With the fifteen wavelet scales used in this paper
we obtain 680 cubic statistics.
Expectation values for each of the 680 are computed by averaging over 300
simulations, after which we compute the change-of-basis matrix $V_{nI}$
using~\eqref{eq:vni}.
We then compute a wavelet map of the real 7-year WMAP data, mask it, and extract cubic
statistics $\tilde{W}_{I}$ in the same way.

In order to compute the $680 \times 680$ covariance
matrix $C_{I J}$, we evaluate the expectation value
$C_{IJ}=\langle \tilde{W}^G_I \tilde{W}^G_J \rangle
- \langle \tilde{W}^G_I \rangle\langle \tilde{W}^G_J \rangle$
over $3 \times 10^4$ simulations.
The wavelet estimator \eqref{eq:fnlWavelet2}
requires the inverse matrix $C_{IJ}^{-1}$,
and once this has been obtained
the quantities
$\barbeta_n^Q$ and $\overline{\gamma}_{nm}$
can be computed.

For each model under consideration we may obtain an estimate for the
$\fnl$ parameter, $\fnlhat^b$, and its expected (Fisher) variance,
$\sigma^2_F(\fnlhat^b)$, using~\eqref{eq:fnlWavelet2} and~\eqref{eq:varfnl}, respectively.
We also compute the variance of $\fnlhat^b$ from a suite of 100 simulations.
Irrespective of whether the linear term~\eqref{eq:waveletlinear} is subtracted,
we recover the expected Fisher variance to high accuracy.%
	\footnote{We replicate the result of \cite{Donzelli2012},
	finding that the variance
	\emph{without}
	accounting for the linear term is within $\lesssim 2\%$ of the variance
	when this contribution is included.}

\subsection{Validation procedure}

\begin{figure}
\centering
\begin{tabular}{c }
\includegraphics[width=0.8\linewidth] {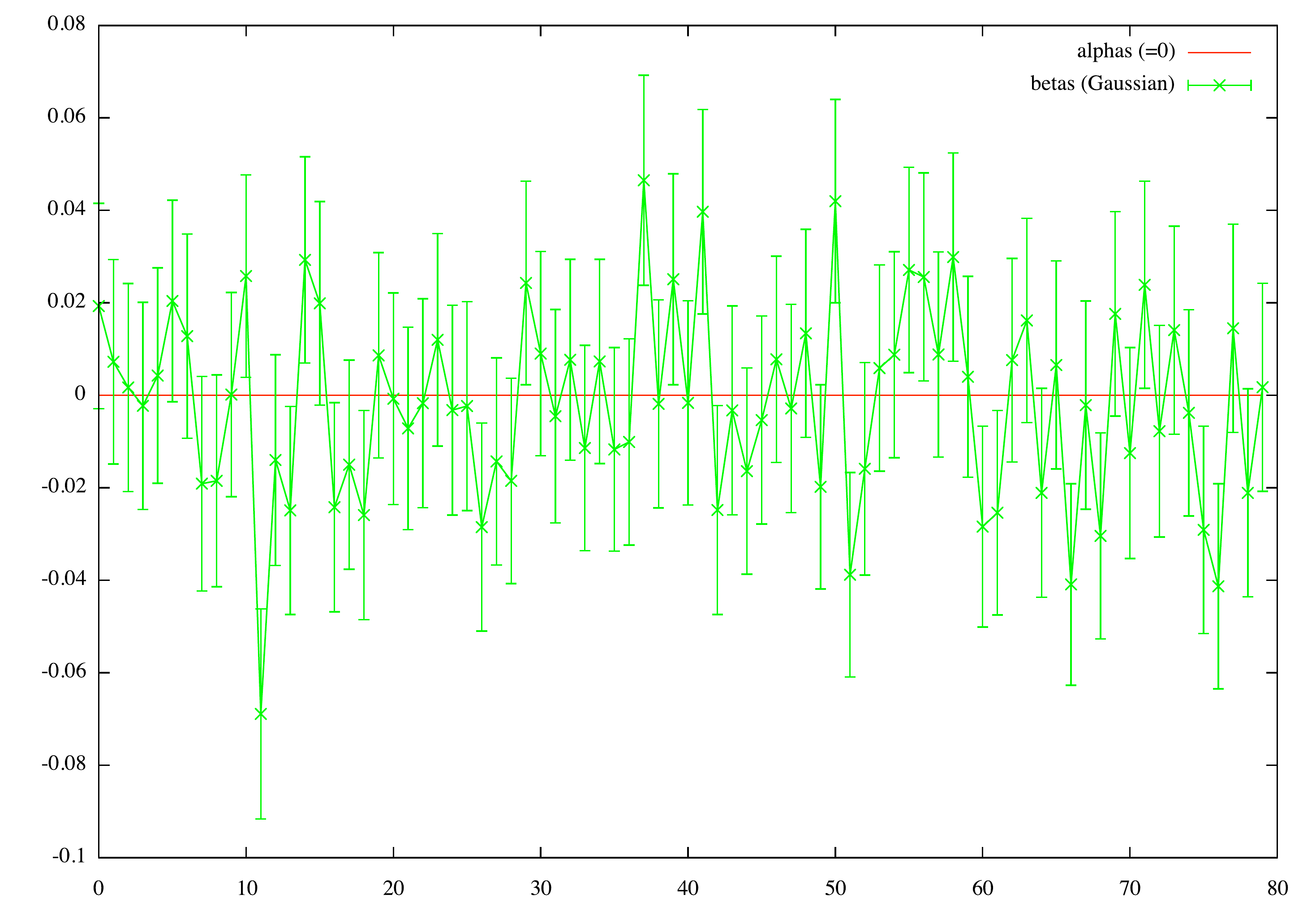}
%&
 %%\includegraphics[width=0.5\linewidth] {/Users/dr200/Desktop/PlotOfMasks/Bispectrum_2D_plots/shape_triangle_local.eps}\\
 % & (b)
\end{tabular}
\caption{Mean of $\langle \barbeta_n^R \rangle$ from $1000$ Gaussian simulations.
To verify consistency of each mode with zero we plot
the standard error of the mean (ie. the standard error divided by the square root
of the number of simulations). Every mode is consistent with zero within two
standard errors of the mean.}
\label{fig:stderr}
\end{figure}

\para{Gaussian validation.}
To ensure that our implementation is unbiased, we perform $4000$ Gaussian simulations.
A critical `sanity' check, described by Eq.~\eqref{eq:sanity},
is to verify that the mean value $\langle \barbeta^R_n \rangle$
for each $n$ is consistent with zero, within the standard error of the mean.
(This is equal to the standard deviation divided by the square root of the
number of simulations.)
We plot the results in Fig.~\ref{fig:stderr}, from which we conclude
that our methodology successfully passes this test.
We have evaluated
the wavelet-based estimator~\eqref{eq:fnlconst} for these simulations with the
result
\begin{equation}
	\langle \fnlhat \rangle = -0.3\pm 23.6. %3\pm 23.5.
\end{equation}
Note 
that the standard deviation $23.6$ recovers the Fisher value
obtained from~\eqref{eq:varfnl},
$\sigma_F(\hat{f}_{\rm{NL}})=23.6$, precisely.
We have verified that neglecting the linear term~\eqref{eq:waveletlinear} 
results in a minimal ($\lesssim 2\%$) difference in the standard deviation.

\begin{figure}
\centering
\begin{tabular}{c }
\includegraphics[width=0.8\linewidth] {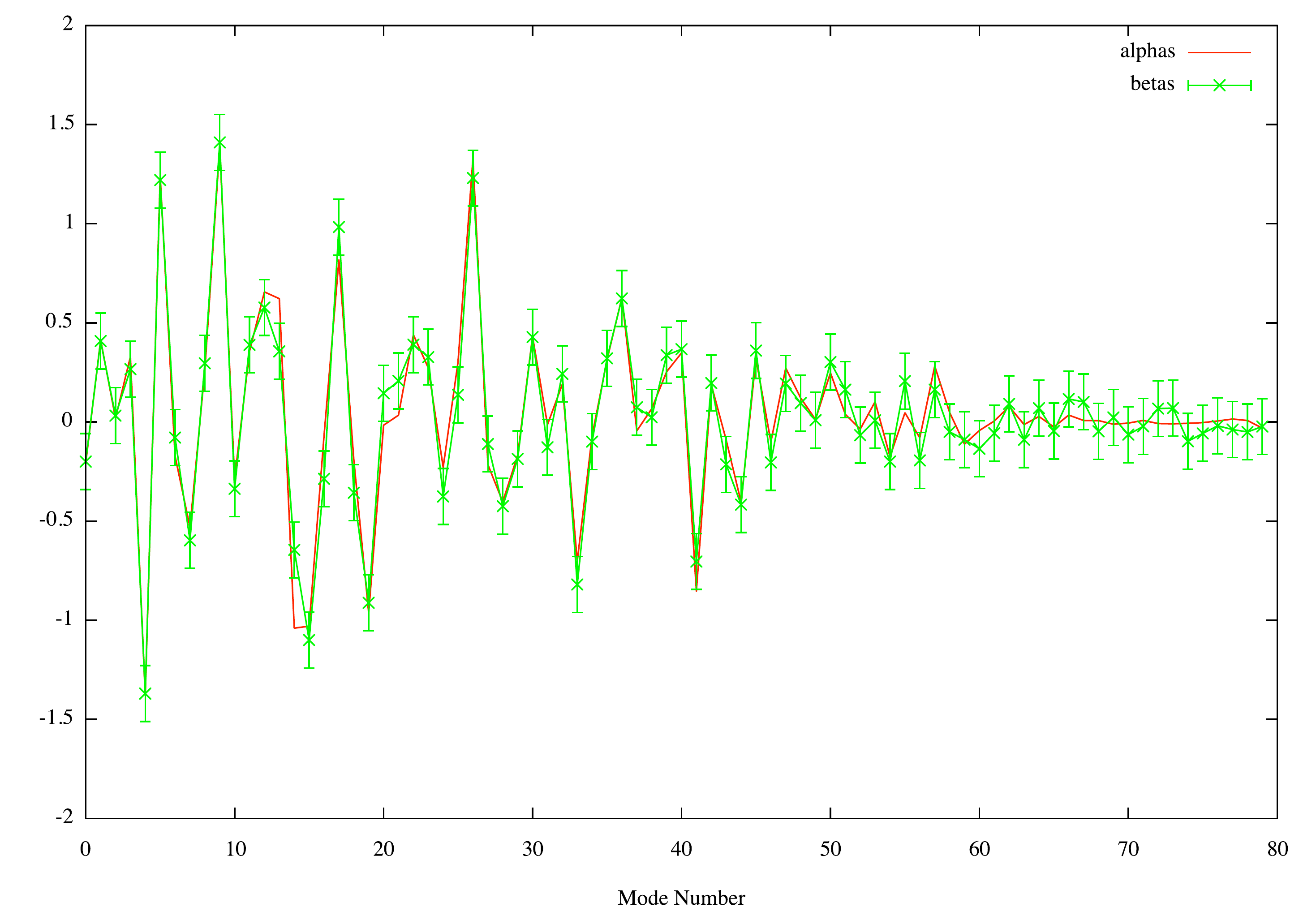}
%&
 %%\includegraphics[width=0.5\linewidth] {/Users/dr200/Desktop/PlotOfMasks/Bispectrum_2D_plots/shape_triangle_local.eps}\\
 % & (b)
\end{tabular}
\caption{Comparison of the modal coefficients for the local shape, $\baralpha_n^R$,
to the observed coefficients, $\barbeta_n^R$.
The observed coefficients are computed by simulating $100$ local maps with
$\fnl^{\text{loc}}=100$ as described in
Ref.~\cite{Hanson:2009kg}. We plot the mean of these modes and the
standard error of the mean.
The `sanity' check \eqref{eq:sanity} is satisfied.}
\label{fig:stderrLoc}
\end{figure}

\para{Non-Gaussian validation.}
We produce
$100$ local simulations with $\fnl^{\text{loc}}=100$, using the method described
in \cite{Hanson:2009kg}. For each of these simulations we extract the observed
modes, $\barbeta_n^R$. The critical sanity check \eqref{eq:sanity}
can now be carried out by comparing the theoretical expectation $\baralpha_n^R$
to each observed mode. In Figure \ref{fig:stderrLoc} we show that this test is again
satisfied within the error bars. The recovered value of $\fnl$ is found to be
\begin{equation}
	\langle \fnlhat \rangle = 99.9\pm 2.5\,,
\end{equation}
where the error bar quoted in this case is the standard error of the mean.

\section{7-year WMAP constraints}\label{sec:wmapconstraints}
In this Section we
use the methodology described in Section~\ref{sec:wmaptest} to
obtain constraints on a selection of nearly scale-invariant models.
To date,
most bispectrum analyses
have considered the local and equilateral models, owing to their physical significance
and computational simplicity.
We extend these to include the constant, orthogonal and flattened templates.
In Ref.~\cite{FLS10} these models were studied using modal methods and
a bispectrum-based analysis.
However, the wavelet-based
approach adopted here yields a
treatment which is closer to optimal.
For example, in the case of local non-Gaussianity the error bar
is reduced from $\Delta \fnl=27.6$ to $\Delta \fnl=23.6$. In Table~\ref{table:comparisons} we compare our results with those of~\cite{FLS10}, highlighting the improvement in optimality achievable with the approach adopted in this paper. The comparison highlights the significant improvement over a standard modal-based approach, which, thus far, has been constrained in its scope, to using the same basis for the estimator as that used for the decomposition of the shape. This necessity may be avoided by employing the change of basis matrix, \eqref{eq:vni}, with the modal technique used for the shape decomposition and wavelets being employed for extraction of the data in this work.
\begin{table}[htp]
\begin{center}
	\heavyrulewidth=.08em
	\lightrulewidth=.05em
	\cmidrulewidth=.03em
	\belowrulesep=.65ex
	\belowbottomsep=0pt
	\aboverulesep=.4ex
	\abovetopsep=0pt
	\cmidrulesep=\doublerulesep
	\cmidrulekern=.5em
	\defaultaddspace=.5em
	\renewcommand{\arraystretch}{1.6}

	\begin{tabular}{SSS}
		
		\toprule
	
		\multicolumn{1}{c}{Shape} & \multicolumn{1}{c}{Current Paper}  & \multicolumn{1}{c}{FLS~\cite{FLS10}}\\
		\toprule
		{\rm{Local}} & \,\,\,38.4\pm 23.6& 20.3\pm 27.6\\
		{\rm{Constant}} & -10.1\pm 60.6& 30.5\pm 95.9\\
		{\rm{Equilateral}} & -119.2\pm 123.6& \quad 1.9\pm 127.4\\
		{\rm{DBI}} & -50.1\pm 104.2& \,\,\,17.1\pm 121.8\\
		{\rm{Orthogonal}} & -173.2\pm 101.5& -51.4\pm 103.8\\
		{\rm{Flattened}}  & \quad\,6.6\pm 10.4&\quad 3.0\pm 10.9\\
		
		\bottomrule
	\end{tabular}
\end{center}
\caption{Comparison of the WMAP$7$ constraints found in this work with those of Fergusson, Liguori and Shellard~\cite{FLS10}. The improvement is achieved by employing a more optimal estimation method, allowed due to the freedom in making a different choice of basis to extract the data from that used to decomposed the theoretical shape.\label{table:comparisons}}
\end{table}

\subsection{Local model}
The local model is defined by a
series expansion of the primordial gravitation potential, $\Phi$,
in powers of an exactly Gaussian potential, $\Phi_G$.
It gives \cite{Salopek}
\begin{equation}
	\Phi(\bx)=\Phi_G(\bx)+\fnl(\Phi_G(\bx)^2-\langle \Phi_G(\bx)^2\rangle)\,.
\end{equation}
Neglecting scale dependence, the local model is an accurate match
for the non-Gaussian contribution generated by
interactions which operate on superhorizon scales.
It generates a bispectrum given by \eqref{eq:locmodel}. 
Examples of scenarios that may produce appreciable
local non-Gaussianity include multifield inflation and curvaton scenarios.
By construction the shape $S_{\Phi}^{\text{(loc)}}=1$ for this model.
For a comprehensive review, see Chen \cite{Chen2010} and references therein.

\begin{figure}
\centering
\begin{tabular}{c}
\includegraphics[width=0.48\linewidth] {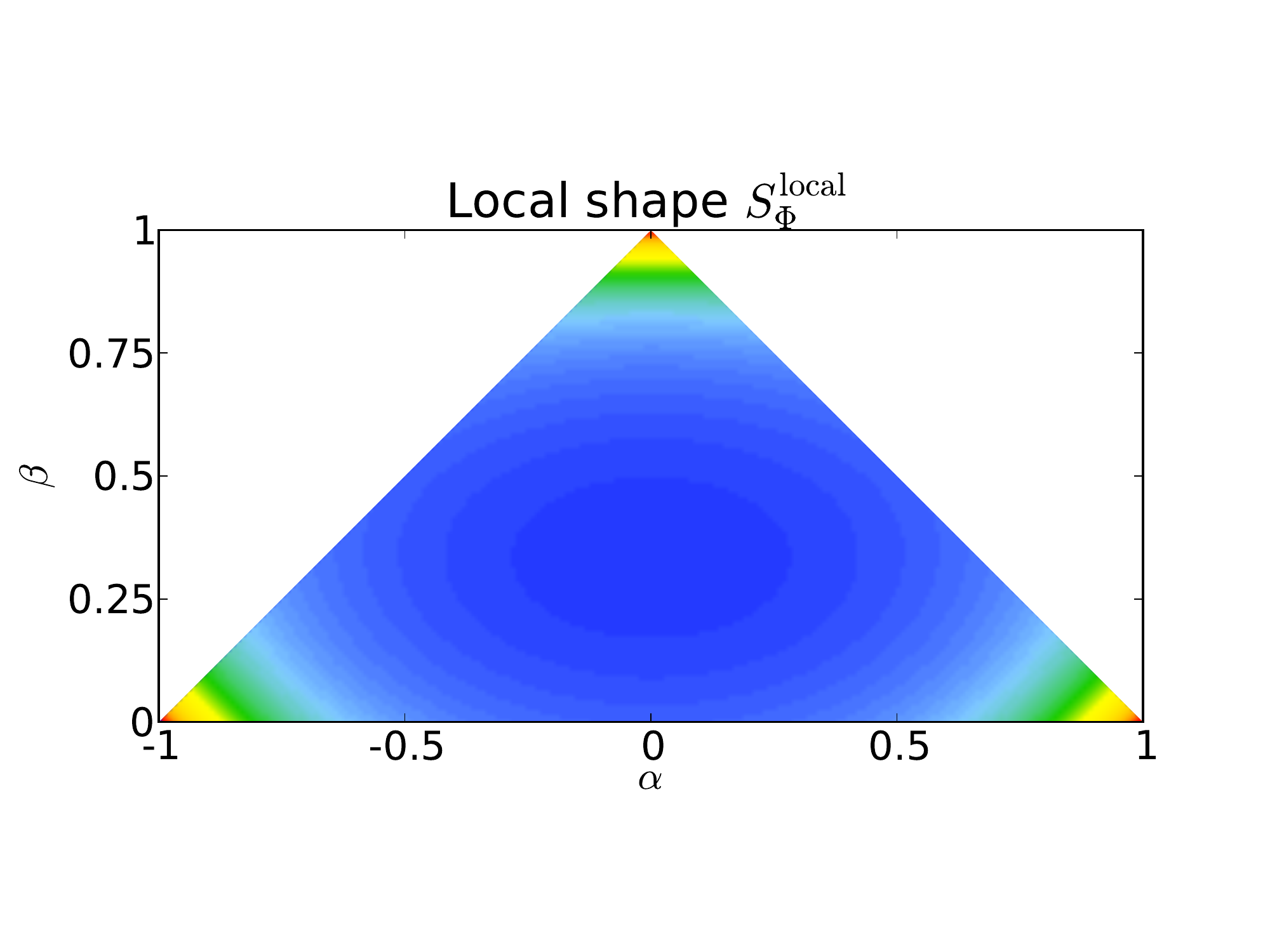}
%&
 %%\includegraphics[width=0.5\linewidth] {/Users/dr200/Desktop/PlotOfMasks/Bispectrum_2D_plots/shape_triangle_local.eps}\\
 % & (b)
\end{tabular}
\caption{Canonical shape function \eqref{eq:shapecanon} for
the primordial local bispectrum. Due to the (almost) scale invariance of the
shape it is only necessary to plot a particular slice. The
parametrization $(\alpha,\beta)$ for each slice is chosen as described in
Ref.~\cite{FLS1}. In particular,
$k_1/k_t= (1+\alpha+\beta)/4$,
$k_2/k_t= (1-\alpha+\beta)/4$,
$k_3/k_t= (1-\beta)/2$,
where $k_t=\sum_i k_i$.}
\label{fig:loc}
\end{figure}

In Figure \ref{fig:loc} we plot the primordial
bispectrum shape.
The dominant signal occurs
in the corners of the triangle, corresponding to `squeezed' configurations
where one momentum is much smaller than the other two.
When transferred to the CMB bispectrum,
this (almost) scale-invariant shape is redistributed, resulting in the presence of peaks in the three-dimensional CMB bispectrum. However, the dominant signal remains along the edges of the tetrahedral domain, i.e. where one $l$ is much smaller than the other
two.

\begin{figure}
\centering
\subfloat[][Comparison of the modal coefficients of the local model against the
recovered modes of the WMAP bispectrum. The coefficients $\baralpha_n^R$
are normalised to aid comparison.]{\includegraphics[width=0.7\linewidth] {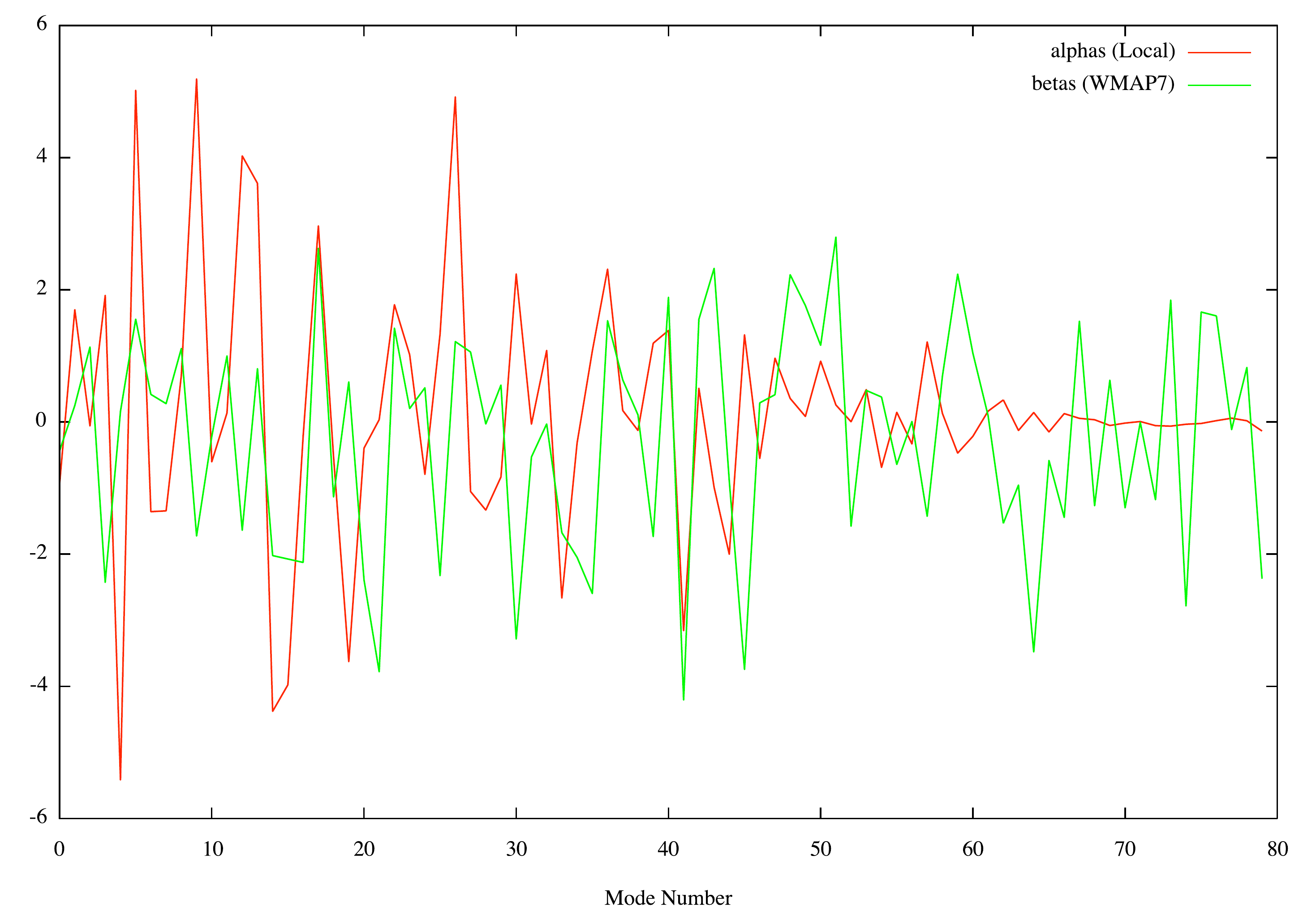}}

\subfloat[][Cumulative contribution of each mode to the amplitude $\fnl$. We plot the partial sum $\sum_{n=0}^{N_{\text{max}}} \baralpha_n^R \barbeta_n^R / \sum_{n=0}^{79} (\baralpha_n^R)^2$ against the maximum mode number $N_{\text{max}}$. Despite the slow convergence of the WMAP signal,
the estimator for $\fnl$ has converged.]{\includegraphics[width=0.7\linewidth] {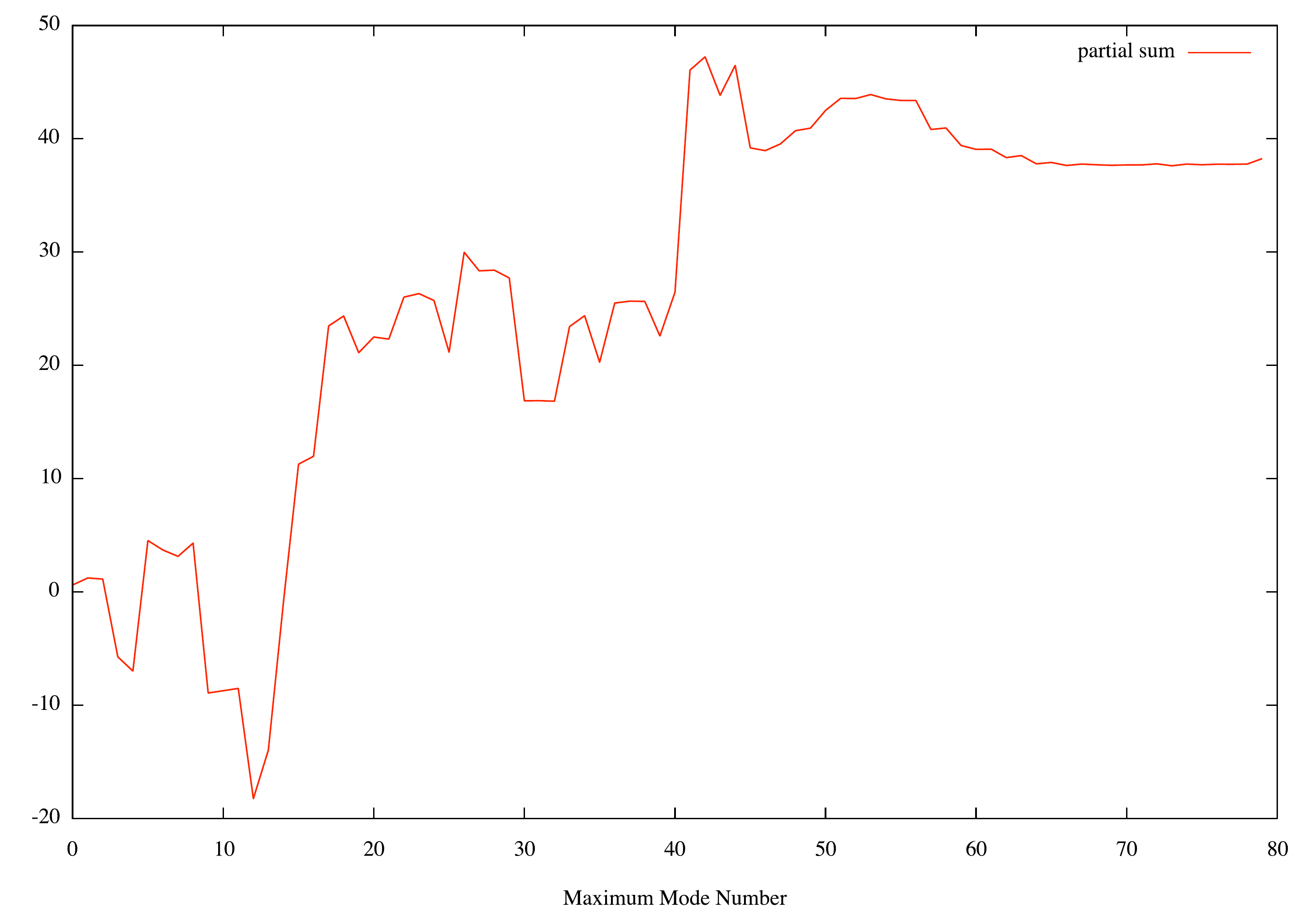}}

\caption{\label{fig:LocvsWMAP}Modal comparison for local model.}
\end{figure}

In Figure \ref{fig:LocvsWMAP}, we compare the modal coefficients for the local model
against the coefficients reconstructed from 7-year WMAP data.
We also plot the cumulative sum $\sum_{n=0}^{N_{\text{max}}} \baralpha_n^R
\barbeta_n^R / \sum_{n=0}^{79} (\baralpha_n^R)^2$
to establish that the estimator does converge with $80$ modes.
Our final constraint on the amplitude of a local-type bispectrum
is
\begin{equation}
	\fnl^{\text{loc}}= 38.4 \pm 23.6
	\quad \text{or} \quad
	\Fnl^{\text{loc}}= 38.4 \pm 23.6 .
\end{equation}
This result is competitive with
the outcome of other wavelet-based analyses.
For example, Donzelli et al. \cite{Donzelli2012} quoted
the constraint $\fnl^{\text{loc}} = 37.5\pm 22.3$. The small difference in
our results may be explained by the use of a slightly different masking procedure,
and
(perhaps more importantly) because their analysis used data up
to $\lmax = 1500$.

\subsection{Constant model}
The constant model
is simply $S_{\Phi}(k_1,k_2,k_3)=1$.
It is interesting because it produces a CMB bispectrum due completely to the transfer
functions.
One possible microphysical realization may occur
during an epoch of quasi-single field inflation \cite{ChenWang2009}.
Our wavelet-based estimator yields the
constraint
\begin{equation}
	\fnl^{\text{const}}= -10.1\pm 60.6% 50 \pm 93.8
	\quad \text{or} \quad
	\Fnl^{\text{const}}= - 3.9 \pm 23.6 .
\end{equation}

\subsection{Equilateral and DBI models}
Interactions which operate on superhorizon scales produce a local-shape
bispectrum
because causality requires the interaction to consist of a long-wavelength modulation
of the background experienced by the short modes.
This correlation between long and short modes is maximized
in the squeezed limit.

In comparison, interactions which dominate on subhorizon scales typically
produce no signal in the bispectrum, because subhorizon modes fluctuate incoherently
and average to zero.
An exception,
where the initial state is non-empty,
will be considered in Section~\ref{subsec:flattened} below.
Neglecting that possibility,
significant effects can be produced only near the epoch of horizon exit,
where the fluctuations are beginning to behave coherently.
In canonical slow-roll, single-field inflation the interference between
horizon-scale fluctuations does not generate significant non-Gaussianity.
However, with non-standard kinetic terms the amplitude of these effects
may be enhanced~\cite{AlishahihaSilversteinTong2004,ChenetAl2007,0605045}.
Examples include DBI inflation and $k$-inflation.

\begin{figure}
\mbox{}
\hfill
\subfloat[][DBI shape]{\includegraphics[width=0.48\linewidth] {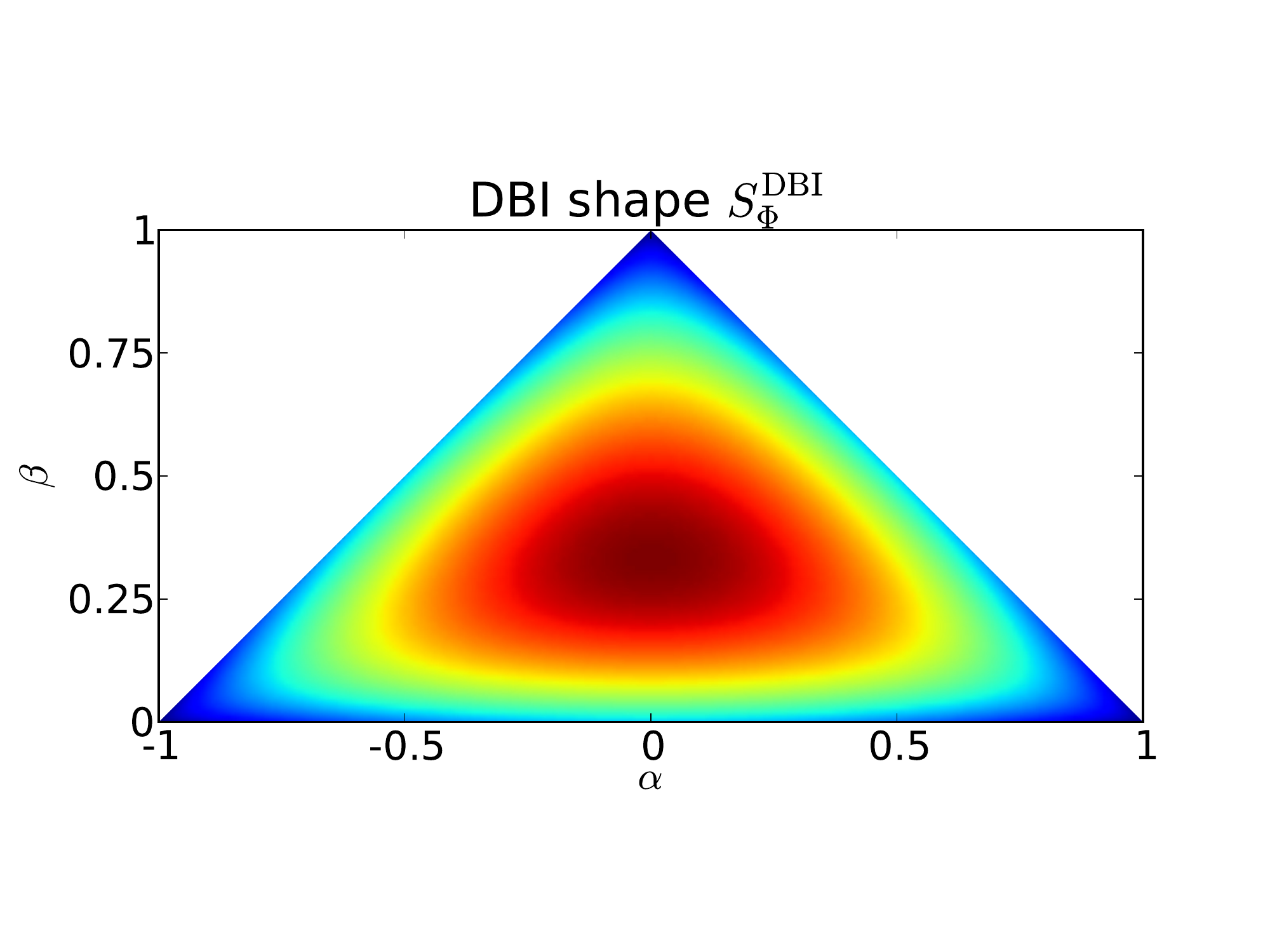}}
\hfill
\subfloat[][Equilateral shape]{\includegraphics[width=0.48\linewidth] {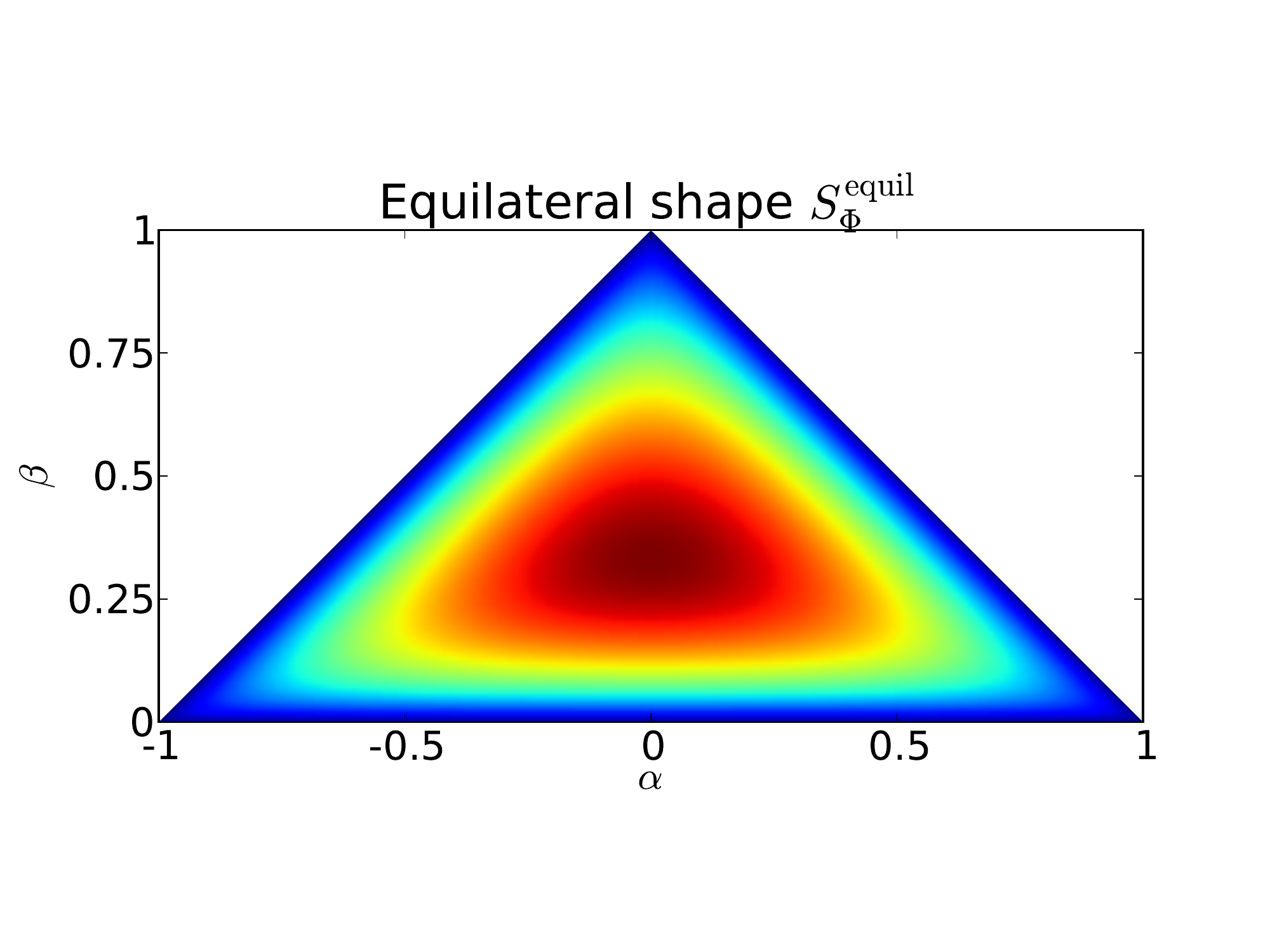}}
\hfill
\par

\caption{Canonical shape function, \eqref{eq:shapecanon}, for the primordial DBI
bispectrum and equilateral bispectrum.
The high degree of correlation between the two shapes, apparent from these plots,
is $98\%$.}
\label{fig:equil}
\end{figure}

The equilateral template is a separable approximation to the bispectrum produced by
such models. It produces strong correlations for roughly equal $k$
because it is dominated by interference effects between wavenumbers which
leave the horizon nearly simultaneously.
We plot the shape function for the DBI model and the equilateral template
in Fig.~\ref{fig:equil}.
They correspond to
\begin{align}
	B_{\Phi}^{\text{DBI}}
	& =
	\frac{1}{(k_1 k_2 k_3)^3 (\sum_i k_i)^2}
	\bigg(
		\sum_i k_i^5
		+ \sum_{i\neq j}(2 k_i^4 k_j-3 k_i^3 k_j^2)
		+ \sum_{i\neq j\neq l}(k_i^3 k_j k_l - 4 k_i^2 k_j^2 k_l)
	\bigg)\,,
	\\
	B_{\Phi}^{\text{eq}}
	& =
	6\bigg(
		- \Big[ P_{\Phi}(k_1)P_{\Phi}(k_2)+\text{2 perms} \Big]
		- 2 \Big[ P_{\Phi}(k_1)P_{\Phi}(k_2)P_{\Phi}(k_3) \Big]^{2/3}
	\nonumber\\
	& \hspace{9mm}
		+ \Big[
			P_{\Phi}^{1/3}(k_1)P_{\Phi}^{2/3}(k_2)P_{\Phi}(k_3) + \text{5 perms}
		\Big]
	\bigg) \,.
\end{align}
Using~\eqref{eq:correlation} it can be
shown that correlation between these shapes is $98\%$.
The wavelet-based estimator gives the constraints
\begin{align}
	\fnl^{\text{DBI}}  = \,\,\,\ -50.1\pm 104.2\quad & \text{or} \quad
	\Fnl^{\text{DBI}}  = -11.4\pm 23.6\,,
	\\
	\fnl^{\text{eq}}  = -119.2\pm 123.6 \quad & \text{or} \quad\,\,\,
	\Fnl^{\text{eq}}  = -22.8\pm 23.6 \,.
\end{align}

A variety of other bispectra
dominated by interference effects near horizon-crossing,
including the case of ghost inflation,
were considered in Ref.~\cite{FLS10}.
All these models are highly correlated with the equilateral template
and produce similar constraints.

\subsection{Orthogonal Model}
The orthogonal shape
is a linear combination of the constant and equilateral shapes, corresponding
to
$S_{\Phi}^{\text{orth}} \propto S_{\Phi}^{\text{eq}}
- (2/3) S_{\Phi}^{\text{const}}$
\cite{SmithetAl2009,MeerburgVanDerSchaarCorasaniti2009}.
It is roughly orthogonal to both the equilateral and local shapes.
A separable template for its bispectrum is
\begin{align}
	B_{\Phi}^{\text{orth}}
	& =
	6\bigg(
		3 \Big[
			P_{\Phi}^{1/3}(k_1)P_{\Phi}^{2/3}(k_2)P_{\Phi}(k_3) + \text{5 perms}
		\Big]
		- 8\Big[ P_{\Phi}(k_1)P_{\Phi}(k_2)P_{\Phi}(k_3) \Big]^{2/3}
		- \frac{3}{2}B_{\Phi}^{\text{loc}}
	\bigg)\,.
\end{align}
It has been shown to arise in the DBI--Galileon model
by Renaux-Petel~\cite{Seb2011}; see also Ref.~\cite{RaquelDavid1}.
We find the following constraint
\begin{equation}
	\fnl^{\text{orth}} = -173.2\pm 101.5  \quad \text{or} \quad%-56.4 \pm 97.2 \quad \text{or} \quad
	\Fnl^{\text{orth}} = -40.3\pm 23.6 \,.
\end{equation}
We note the consistency of this result with the
constraint $\fnl^{\text{orth}}=-159.8\pm 115.1$ obtained
by Curto et al.~\cite{Curto2011}.

\subsection{Flattened Model}
\label{subsec:flattened}
If the initial state for fluctuations is not empty then it is possible to produce
a bispectrum describing maximum correlation for `flattened'
configurations where, eg., $k_1 \approx k_2 + k_3$.
The correlation arises because, with a nontrivial initial state, it is possible
to find a `negative' energy fluctuation with time dependence $\sim \e{-\im k_1 t}$
which interacts coherently with two positive energy fluctuations
with time dependence $\sim \e{+\im k_2 t}$, $\e{+\im k_3 t}$.
When $k_1 \approx k_2 + k_3$ the interaction is coherent
over arbitrarily long times
and does not average to zero in the subhorizon era.

Holman \& Tolley studied a model in which the bispectrum produced by this effect
was~\cite{HolmanTolley2008}
\begin{equation}
	B_{\Phi}^{\text{flat}}
	=
	\frac{6}{39(k_1 k_2 k_3)^2}
	\Bigg[
		\left(
			\frac{k_1^2+k_2^2-k_3^2}{k_2 k_3} + \text{2 perms}
		\right)
		+ 12
		+ 8 \left(
			\frac{k_1 k_2+ k_1 k_3 -k_2 k_3}{(k_2+k_3-k_1)^2} + \text{2 perms}
		\right)
	\Bigg]\,.
	\label{eq:flattened-bispectrum}
\end{equation}
This is not separable.
Its analysis is computationally intensive without a method such as modal decomposition,
although
alternatives approaches exist such as the use of Schwinger parameters~\cite{0612571}.

Eq.~\eqref{eq:flattened-bispectrum} diverges in the flattened
limit because the interaction
continues
over arbitrarily long times, and therefore becomes sensitive
to whatever physics was operative throughout the inflationary era.
By comparison, inflationary predictions using a vacuum
initial state decouple from this unknown
high-energy physics.
To handle the divergence we parametrize our ignorance
of the relevant physics
using a cutoff~\cite{FLS1},
setting the bispectrum to zero for $k_1 + k_2 - k_3 < Z k_t$
(or its permutations), where
the perimeter
$k_t = k_1 + k_2 + k_3$
was defined below Eq.~\eqref{eq:triangle-condition}.
In this paper we take $Z = 0.03$ as a fiducial value, although in a dedicated analysis
$Z$ should be allowed to float. Furthermore, we smoothen the shape near the edges by employing a low pass (Gaussian) filter. 
With this choice,
our
constraints on the flattened model from the 7-year WMAP data are
\begin{equation}
	\fnl^{\text{flat}} = 6.6\pm 10.4%-2.7 \pm 7.5
	\quad \text{or} \quad
	\Fnl^{\text{flat}} = 15 \pm 23.6 \,.
\end{equation}

\section{Conclusions}\label{sec:conclusions}
In this paper we have developed a framework which combines partial-wave or
`modal' techniques with wavelet-based estimators for the CMB bispectrum.

Wavelet-based techniques
are particularly efficient for CMB analysis because they
take advantage of simultaneous localization
on the temperature map in
both scale and position.
However, it is not straightforward to build a wavelet-based estimator for the
amplitude of an arbitrary primordial bispectrum $B_{\Phi}$. Combining
the wavelet-based methodology with the decomposition of
$B_{\Phi}$ into a basis of partial-waves enables an efficient
analysis---whether $B_{\Phi}$ is separable or not, provided only that it is a
relatively smooth function of wavenumber.

Our framework has other advantages.
Optimal bispectrum-based estimators are typically hampered by the need to invert
a pixel-by-pixel covariance matrix
of size $\sim 10^6 \times 10^6$.
Because the wavelet-by-wavelet covariance matrix is typically much smaller,
of order $\sim 10^3 \times 10^3$,
the problem is numerically much more tractable.
Despite this large gain in numerical efficiency there is comparatively little
trade-off in optimality. Indeed, our final error bars are quite close to those achieved
by the optimal pixel-by-pixel approach.
In addition, due to their localization properties,
wavelets have been shown to allow for accurate analysis of point-sources,
foregrounds, and other systematics~\cite{Curto0807,McEwen0704}.
Hence, integrating the partial-wave methodology with a wavelet-based analysis
has the potential to reproduce the successes of both.
The work presented in this paper generalizes the scope of wavelet-based estimation
to allow for analysis of arbitrary primordial bispectra.

We have implemented our methodology for the 7-year WMAP data.
Our constraints are competitive (to within $\sim 5-10\%$) with comparable
constraints published elsewhere, and represent an improvement of up to $\sim 15\%$
in comparison with the bispectrum-based modal estimator of Ref.~\cite{FLS10}.
In any case there is much to be gained from implementing different
estimators: they will be sensitive to different combinations of the data,
including the underlying systematics.

In future work, we intend to study the efficacy of the method in more detail
and pursue an extension to the trispectrum. (See also
Refs.~\cite{RSF10,FRS2}.)
Although our constraints show that the 7-year WMAP data are consistent with
Gaussianity, we will shortly be presented with an improved data set from
\emph{Planck}. We hope that the techniques described in this paper can help
to correctly categorize the source of any signal---whether
of primordial origin, or a foreground.

\section*{Acknowledgements}
It is a pleasure to thank
Antony Lewis, Andrew Liddle, Raquel Ribeiro and
S\'{e}bastien Renaux-Petel for helpful discussions.
We thank Mateja Gosenca for generating some of the images
used in this paper.
DMR acknowledges a long collaboration with James Fergusson and
Paul Shellard in developing many aspects of the modal methodology.

Some numerical presented in this paper were obtained using
the COSMOS supercomputer,
which is funded by STFC, HEFCE and SGI.
Other
numerical computations were carried out on the Sciama High Performance Compute (HPC)
cluster which is supported by
the ICG, SEPNet and the University of Portsmouth.
We acknowledge support from the Science and Technology
Facilities Council [grant number ST/I000976/1].
DS acknowledges support from the Leverhulme Trust.
The research leading to these results has received funding from
the European Research Council under the European Union's
Seventh Framework Programme (FP/2007--2013) / ERC Grant
Agreement No. [308082].

\bibliography{WaveletsPaper}

\end{document}